**Applicability test for reducing noise on PET dynamic images using phantom applying deep image prior**

Nobuyuki Kudomi[1] and Yukito Maeda[2]

[1]Department of Medical Physics, Faculty of Medicine, Kagawa University, Kagawa, Japan

[2] Department of Clinical Radiology, Kagawa University Hospital, Kagawa, Japan

**Running title:** Denoising PET dynamic image

**Corresponding author**

Nobuyuki Kudomi

Address: Department of Medical Physics, Faculty of Medicine, Kagawa University, Kagawa, Japan

E-mail: kudomi.nobuyuki@kagawa-u.ac.jp

Phone & Fax: +81-87-891-2247

**Abstract**

*Objective* Positron emission tomography (PET) allows imaging of patho-physiological information as a form of rate constants from a dynamic image. The rate constant image(s) may be affected from noise on the dynamic image. We introduced an artificial intelligence technique of deep image prior (DIP) to reduce noise on dynamic images.

*Method* We utilized a phantom filled with $^{18}$F-F$^-$ and $^{11}$C-flumazenil solutions in the main and sub-cylinders, respectively. The phantom was scanned by a Biograph mCT and dynamic images were obtained. DIP was applied to all slices involved in the dynamic images while introducing an index for choosing an optimal epoch with minimize the degree of noise. Then, decay rate images were generated and quantitative accuracy and quality were measured in the images.

*Results* The obtained decay rates on images were not significantly different from those of the reference values. Coefficient of variances (CV) were smaller using DIP-based images than those from original images for the reconstructed and decay rate images, both in the $^{11}$C and $^{18}$F filled regions.

*Conclusion* The present method for choosing a generated image during DIP is feasible for noise reduction in dynamic PET images, and for obtaining rate constant images with less noise.

**Keywords**    PET Image Denoising, · PET · Kinetic analysis, Phantom Study, Deep Image Prior, Dynamic Image

**Introduction**

Positron emission tomography (PET) using radiotracers with quantitative analysis provides physiological information of organs as a form of rate constant(s) [1]. Quantitative accuracy and, in particular, image quality in PET image have been improved in PET devices and reconstruction algorithms [2], with such benefit as the advancement of reconstruction algorithms involving filtered back projection (FBP), and ordered-subsets expectation maximization (OSEM), which in imaging is remarkable in the nuclear medicine field [3]. Several studies have demonstrated and characterized those reconstruction algorithms.

In addition to the development of the reconstruction algorithms, recent improvements have incorporated artificial intelligence (AI) techniques, such as deep image prior (DIP), originally developed by Ulyanov et al, consists of so-called 'U-net' architecture [4]. A characteristics of the net is earlier convergence of structural shapes on images than that of noise. One advantage of DIP is without requiring of amount of supervisor images, namely, we can apply the method only one image, and thus quite practical for the nuclear medicine field. In that field, the incorporation of DIP has resulted in improvement in noise properties [5,6,7].

Some studies have assessed the noise properties in PET and single photon emission tomography (SPECT)-based images with DIP; however, no studies assessing appropriate epoch number during image generating procedure starting from random and ending with noise involving images. Cheng et al [8] assessed the appropriate epoch number in non-nuclear medicine study; however, their implementation did not provide the option of an early choice of an optimal image during the generating phase. For noise reduced parametric images after kinetic analysis, automatic and time saving strategies for denoising all images involved in a PET dynamic image are needed. In the present study, we introduced an index to allow the selection of one optimal generated image during the DIP procedure; denoised dynamic images were generated for dynamic images scanned with a phantom generated by FBP and OSEM algorithms. For feasibility, rate constant images after kinetic analysis were computed for two different decay rate isotopes of $^{18}$F and $^{11}$C, and quantitative accuracy and quality were obtained in those rate constant images. In addition, we tested the degree of mixture around the boundary portion of the image.

**Materials and Methods**

Phantom and PET scan

The scan data used was the same as that reported in our previous study [9]. Briefly, the JSP model phantom (KYOTO KAGAKU CO., LTD, Kyoto, Japan) (Fig. 1) was used, which has a cylindrical shape 200 mm in the inner diameters. The unit consists of seven sub-cylinders. The phantom was filled with $^{18}$F-F$^-$ ion solution in the main component and with $^{11}$C-flumazenil in all sub-cylinder regions, respectively.

The PET scanner used was the Biograph mCT64-4R PET/CT system (Siemens Medical Solutions,

Knoxville, TN, USA). The activity concentration of each solution was 20 kBq/ml and 30 kBq/ml for $^{18}$F and $^{11}$C, respectively, at the start time of the PET scanning.

Image reconstruction

The collected list mode data were sorted to produce dynamic sinogram, with 300 sec × 24 frames and 120 min in total. Images were then reconstructed by the FBP and OSEM bases with or without the incorporation of the time of flight (TOF), and point spread function (PSF) for OSEM, without decay correction, including corrections for dead time, detectors normalization, CT-based attenuation, and scatter [10] using the vendor software programs. We then obtained FBP, FBP with TOF (FBP+TOF), OSEM, and OSEM with TOF and PSF (OSEM +TOF+PSF). For the OSEM procedure, the applied conditions were 3 iterations and 24 subsets; for OSEM+TOF +PSF, the conditions were 3 iterations and 21 subsets. Filtering was not applied during the reconstruction procedure. Filtering was subsequently applied on the reconstructed images with Gaussian filters of 1 and 3 mm, respectively. The reconstructed images consisted of a 400×400×45 matrix, with a pixel size of 2.04 mm × 2.04 mm × 5 mm and with 24 frames. Those images were cropped to a matrix size of 128×128×1 for this study, and the slice 23$^{rd}$ was chosen as we used in our previous study [9].

Denoising Procedure

For obtaining denoised and restructured image, we needed to choose one optimal image generated during the DIP convergence procedure. The required conditions for this are to keep the quantitative accuracy and to minimize the noise degree. To manage those conditions, first, we used pixels set with values higher than that of the threshold set at 10% from the maximum in an input original image. For the quantitative accuracy, we computed the differences of the mean of the pixel values between the original and DIP-generated images. The obtained difference was divided by the mean of the original image, that is, the coefficient of variance (CV), during the DIP procedure, and the required condition were set as < 1% ($CV_{OvsG}$). For minimizing the noise degree, we introduced the $N$-index [11], which allows for an estimation of the degree of noise between two similar but statistically independent images. $N$-index in that previous study was obtained by subtracting those two similar images from one another and computing the standard deviation (SD) on the subtracted image. To measure the $N$-index in the present study, we generated a mean image of the preceding and following frames to the targeted original, and the $N$-indices were computed between the DIP-generated and the mean images. The obtained $N$-indices were divided by the mean of the pixel values in the original image, and the CV was obtained ($CV_{PF}$).

DIP was applied to all images among the reconstructed and filtered dynamic images. The neural network consists with U-net, or encoder-decoder network. A unit in encoder has 2D convolution layer with a set of 3×3 2D filters, rectified linear unit activation, and followed by averaged pooling. The unit repeated 3 times. The decoder network was applied after the encoder network, wax the reverse process of the encoder as mirrored layers from the encoder. The network also features shortcut connections. The total number of epochs was set to $10^4$ and $CV_{OvsG}$ and $CV_{PF}$ were monitored in every 10th epoch. The generated images with the minimum $CV_{PF}$ as well as the $CV_{PF} < 1\%$ were chosen as optimals. Our network was implemented in the Tensorflow (2.5.0) and incorporated with the Keras framework and Python 3.8.6. The computation was performed on a computer using Fedora 32 with one NVIDIA GeForce 3070 GPU. The CUDA library was 11.2 and the cuDNN version was 8.0.5.

Kinetic analysis

To test the effect of reduction of noise in images for kinetic analysis, decay rate images were computed by fitting a single exponential function as described in our previous estimation [9]. Briefly, a tissue curve ($C(t)$) can be expressed as; $C(t) = K_1 \cdot A(t) \otimes e^{-k_2 t}$, where $K_1$ and $k_2$ are forward and backward transfer rate constants from blood to tissue, $A(t)$ is an input function determined to be the delta function, i.e., $A(t) = \delta(0)$, and $\otimes$ denotes the convolution integral. The basis function method (BFM) was applied for estimating those rate constants [12]. The $k_2$ corresponds to decay rate images which were computed for all of the original and DIP noise reduced dynamic images, for 60 min starting at 0 min (first half) and at 60 min (second half).

Data analysis

Two circular regions of interest (ROI) (12.2 mm in diameter) were placed in a slice, one was placed at

the center of the sub-cylinders in the central portion of the $^{11}$C regions and the other was placed at 50

mm left from the center of the phantom in the $^{18}$F region (Fig. 1). CVs on those two ROIs were obtained for the original slices and the DIP generated images ($CV_C$ and $CV_F$ for $^{11}$C and $^{18}$F regions, respectively).

CVs for those ROIs on the decay rate images were also obtained. The means ($\pm$SD) of the decay rate value were extracted from the ROIs for the obtained decay rate images.

Profiles of the cross-sectional distribution of the decay rate images for both the first and second halves were extracted from top to bottom in a vertical direction at the horizontal 128th pixel.

**Results**

The average time required to compute the $10^4$ epochs was 108 s. The epochs which generated images

were chosen and were summarized in Table I. The mean of the epochs for all images was $1929 \pm 2440$

and with a minimum and maximum of 160 and 9800, respectively. The epoch was larger when the size of the filter was larger. No tendency was found in the slope value between the epoch numbers. Those values were from -158 to 42 (epochs)/(frame number) and the mean$\pm$SD was $42 \pm 57$.

The computed CVs ($CV_{PF}$, $CV_{OvsG}$, $CV_C$, and $CV_F$) against the epoch for FBP without filtering for the

22$^{nd}$ frame and for the OSEM+TOF+PSF with the 3 mm filter for the 2$^{nd}$ frame, are shown in Fig 2.

The generated images during the procedure for several epochs are also shown. Those values for all frames for all reconstruction algorithms and filter sizes are presented in the supplemental file (Fig S1 and S2). For the FBP, the $CV_{PF}$ curve initially decreased, reached the minimum, and then increased. For the $CV_{OvsG}$ curve, the curve gradually approached zero from the beginning and subsequently fluctuated around zero. For the images, the structure of the phantom appeared to be gradually generated and noise components were also generated as the epoch increased. For OSEM+TOF+FBP, the $CV_{PF}$ curve continuously decreased. These characteristics are similar to those of other cases for other algorisms and filter sizes.

Fig 3 shows the relationships of $CV_C$ and $CV_F$ between before and after DIP denoising for reconstructed and decay rate images. The dashed line in black shows $y = x$. Noise degree was similar on that line, and the degree was reduced in the area below the line. Most of plots were placed below the line for both reconstructed and decay rate images.

Fig 4 shows the decay rate images, and the decay rate values for the ROIs on the $^{18}$F and $^{11}$C filled regions are summarized in Table II. The quality of the images appeared to improve in the DIP-based images. The rate values in the $^{18}$F region were similar between those from the original and those from the DIP-generated images and close to a known physical decay constant. SDs for those values were always reduced. In the $^{11}$C region, the decay rate was similar between the original and that of the DIP except in the second half for FBP, close to the reference value for most cases from the first half, but underestimated from the second half for OSEM with the 3 mm filter in the DIP images.

Profiles of the cross-sectional distribution in the estimated decay rate images are shown in Fig. 5. In the profile structure, the noise degree reduction can be seen for DIP-based images. At the boundary portion, the decay rate gradually changed from $^{11}$C to $^{18}$F decay rates depending on the ratio of the distance from the regions filled with those solutions, and the degree of change appeared to be decreased from the images without filtering and increased with the size of filter in the larger ones. A reduction of that degree can be seen for the DIP-based images. In addition, some degree of recovery of the underestimation in the decay rate can be seen for the OSEM based images in the second half.

**Discussion**

In the present study, we performed noise reduction on dynamic images by means of a slice-by-slice wise manner while incorporating DIP, and decay rate images for two different decay rate isotopes of $^{18}$F ($\lambda$=0.00631 min$^{-1}$) and $^{11}$C ($\lambda$=0.0341 min$^{-1}$) were obtained from those dynamic images. We assumed that the physical characteristic of decay constant corresponds to a physiological kinetic parameter, in particular the decay constant of $^{11}$C is not very far to physiological transfer rate constants from brain tissue back to blood, i.e., $k_2$, such as 0.13±0.07 min$^{-1}$ and 0.11±0.04 min$^{-1}$ for $^{18}$F-FDG in brain gray and white matter regions [13], respectively, ranged from 0.099 to 0.21 min$^{-1}$ for $^{18}$F-flumazenil in brain cortical regions [14], and 0.10±0.07 min$^{-1}$ for $^{18}$F-FLT in glioma in brain [15]. We also assumed that the present phantom simulated objects to be scanned such as organs with changing distribution of tracer though structure of the phantom was simple, which enables estimating CV directly. The obtained decay rates after noise reduction were essentially not different from the reference rate constants for both of the used isotopes, suggesting that quantitative accuracy would not deteriorate after the DIP denoising procedure. For quality in the decay rate images, CVs for $^{11}$C and $^{18}$F regions were smaller in DIP-based images than those from the original based images. The present test study with simple but having structure would be essential for denoising on an image by mean of DIP before applying to clinical PET dynamic images.

Many methods have developed for denoising dynamic images in nuclear medicine filed. Before appearing the deep learning-based methods in the field, the Gaussian filtering method was widely applied as in the present study. Quality of obtained images after filtering were improved, but a drawback is enhancement of degree of mixture of regions, particularly around boundary part with different rate constants in an organ. After the deep learning-based methods are introduced in the field, one of a possible technique for denoising would be applying U-net architecture [16-18]. The method requires amount of data, such as hundreds, and in particular supervisor images, meaning that amount of pair of noise involved and less noise images are required. For reducing noise on each image consisting in a dynamic image, less noise supervised image might not be practical due to short scan duration and non-reproducible property. The DIP method, as applied in the present study, is a practical approach for this purpose, namely only single image involving noise can be applied. The characteristics of that net is earlier convergence of structural shapes on images followed by noise component. In other words, structural shape is easy to reproduce in neural networks, but random noise is not.

The present study aimed to investigate the improvement of parametric images regarding image quality, without deteriorating quantitative accuracy. The decay rates obtained after noise reduction were similar to those seen in the original images. Thus, quantitative accuracy did not deteriorate. In FBP without a filter for the $2^{nd}$ half, the image still showed significant noise, and the accuracy did not appear improve. In the $^{11}C$ region, an underestimation in the decay rates for the DIP-based images was still seen, but was closer to the reference rates in some images than those for the original based images. The underestimation in the DIP-based images would be due to simultaneous reconstruction of the noise component as a structure in applied images. Regarding quality, the degree of noise reduction was greater when the CV was more than 15% in the original based images, and the degree was similar when the CV was less than that level. This suggests that the present approach would be of value when generating parametric images with kinetic analysis which could potentially have significant noise at a level >15%.

Regarding the DIP procedure, for FBP without filtering, the monitored $CV_{PF}$ was decreased from the beginning, reached a minimum, and subsequently increased. The characteristics were similar for images without filtering and for images with a larger noise degree. In the minimum phase, $CV_{OvsG}$ was not always less than 1%, i.e., the reconstructed image was not quantitatively acceptable, but the acceptable image was generated when the $CV_{PF}$ was in the increasing phase. This is due to the image at the $CV_{PF}$ minimum being very uniform in the generated images with that of the shown images above the CV curves (Fig 2). Thus, the structure in the object was not reconstructed sufficiently at the $CV_{PF}$ minimum phase. After that point, the structure in the images was gradually reconstructed, followed by noise generation. For those CV curves, once $CV_{PF}$ reached the minimum, then it monotonically increased. We can choose one generating image during this phase with the required $CV_{OvsG}$ as < 1%, before noise components are generated. Actually, $CV_{PF}$ fluctuated during the increasing phase, and a possible strategy for stopping the DIP estimation is to continue with hundreds of epochs from the epoch of $CV_{PF}$ at a minimum. For OSEM+TOF+PSF with 3 mm filtering, the monitored $CV_{PF}$ was continuously reduced except in the exceptional phase at approximately 7500 epochs. The characteristics were similar for images with filtering and for image with a smaller noise degree, and chosen generated images would be applied for the following analysis. The present noise reducing procedure would be acceptable or both the above cases. Actually, regarding the $CV_C$ and $CV_F$ curves, when directly measuring the degree of noise on uniform regions, the curves were increased ($CV_C$) or constant with fluctuation ($CV_F$) during the phase in which for the generated image was chosen.

The mean of the epochs in which the generated images were chosen was 1929, and computation time for $10^4$ epochs was 108 s. This data suggests that the computation time for one slice was approximately 20 s, and approximately 8 hours for one dynamic image with 25 frames and 50 slices. More time is necessary with increased image matrix size. The present study also showed that with an increased degree of noise, less epochs were required, suggesting that the time would be shortened to approximately 2 hours. This time still may not be considered as short; however, it would be of value in applying denoising when images after kinetic analysis exhibit significant noise such as in multi-parametric images for reversible tracers with the two tissue compartment model [19]. Further study is needed for the applicability of this technique to clinical images.

The maximum epoch in which the generating images were chosen was 9800 for the 2nd frame by the OSEM+TOF+PSF with 1 mm filtering. The shapes of the $CV_{PF}$ curve was not similar to that of the other cases. In fact, the other cases sometimes showed a stepwise decrease. These results are similar in FBP with 1 mm filtering for the 11th frame and in OSEM without filtering for the 12th frame. The generated images showed that one or some of cylinder(s) was/were not reconstructed sufficiently in the early phases and that a larger $CV_{PF}$ was measured. However, after epochs, the structure of cylinder appeared, and acceptable images were generated. This may have occurred due to introducing random numbers when generating the noise image for the onset as well as in the network architecture. Nevertheless, in this case we obtained denoised images. One option may be to restart the DIP procedure in case in which we did not obtain an optimal image.

The present study investigated the degree of noise reduction. In most cases the results showed that the degree was less than $1/\sqrt{2}$, meaning that, statistically, with the noise degree on an input image reconstructed with half the amount of lines of response (LOR), the DIP-noise reduced image would still have better quality than that of a reconstructed image with the full amount of LOR. Therefore, two images can be reconstructed using both half amounts of the LOR from the list mode data; one could be used as an input for DIP and the other as the $N$-index computation, and vice versa. Then, the mean image of the two DIP-generated images would be further noise reduced. Thus, the present method would be applicable for both dynamic and static purpose images.

In the profile structure at the boundary portion, the decay rate gradually changed between the $^{11}$C to $^{18}$F regions, and the degree of change seemed greater as the size of the filter increased due to the mixture of the decay rate. To avoid the mixture between the different rates, reconstruction without filtering may be better. In that condition, the degree of the mixture was reduced; however, the noise component remained, particularly in the rate images from the 2nd halves for the FBP. For the FBP and OSEM with TOF (and PSF), when a smaller filter of 1 mm was applied, the mixture and noise degrees seemed to decrease in the 2nd half, suggesting that the incorporation of smaller filtering and DIP noise reduction would be practical.

It would be better to reduce noise not on reconstructed images but on directly acquired data such as sinograms, because some of factors are involved such as attenuation correction and filtering. We reconstructed images with commercially available software which may not have any option to accept processed sinograms. Our practical choice was applying the reconstructed image without filtering. Still, we found improvement regarding image quality with keeping quantitative accuracy.

The limitations of the present study were; the study only estimated the decay rate corresponding to the physiologically backward transfer rate, meaning that the accuracy of the forward rates was not directly tested. In addition, the width of the boundary between $^{11}$C and $^{18}$F was not sufficiently thin, i.e., 3 mm to simulate the physiological difference of rate constants at adjacent regions. Finally, the structure of phantom used in this study was simple and not similar to any organs with more complex dynamics, Therefor, the validity of the present method for clinical application remains unclear.

In conclusion, the present study revealed that DIP application can improve image quality in the rate images while keeping the quantitative accuracy, and that the present method may be applicable for denoising PET dynamic images accompanying kinetic analysis. This could lead to the possibility of obtaining parametric images of reasonable quality which to date have not been practical.

Declarations

Acknowledgement

The authors thank Prof Keisuke Matsubara, in Akita Prefectural University, for his kind instruction to introduce DIP code and usage. The authors also thank the staff at Department of Clinical Radiology in our University Hospital and at Department of radiology in our University.

Conflicts of interest/Competing interests

The authors declare no conflict of interest.

Ethics approval

Owing to the nature of this study, the ethical review requirements were waived by our medical ethics review committee.

Informed consent

Owing to the nature of this study, the informed consent was not required.

Authors' contributions

Both authors contributed substantially to the scientific process leading to this manuscript, and participated in the study design and the data analysis. YM prepared activity solutions and acquired the data. NK performed DIP and analyzed data.

Consent for publication

All authors read and approved the final version of the article.

Availability of data and material

All data and additional file are available upon request from the corresponding author.

**References**

[1] Watabe H, Ikoma Y, Kimura Y, Naganawa M, Shidahara M, PET kinetic analysis--compartmental model. Ann Nucl Med 2006; 20: 583-8.

[2] Peng H, Levin CS. Recent developments in PET instrumentation. Curr Pharm Biotechnol. 2010; 11: 555–71.

[3] Tong S, Alessio AM, Kinahan PE. Image reconstruction for PET/CT scanners: past achievements and future challenges. Imaging Med. 2010; 2: 529–45.

[4] Dmitry Ulyanov, Andrea Vedaldi, Victor Lempitsky, Deep Image Prior, Proceedings of the IEEE Conference on Computer Vision and Pattern Recognition, 2018, pp. 9446-54

[5] Fumio Hashimoto, Hiroyuki Ohba, Kibo Ote, Akihiro Kakimoto, Hideo Tsukada and Yasuomi Ouchi, 4D deep image prior: dynamic PET image denoising using an unsupervised four-dimensional branch convolutional neural network. Physics in Medicine & Biology, 2021; 66: 015006

[6] Cheng-Hsun Yang & Hsuan-Ming Huang, Simultaneous Denoising of Dynamic PET Images Based on Deep Image Prior Journal of Digital Imaging. 2022; 35: 834–45

[7] Unsupervised PET logan parametric image estimation using conditional deep image prior. Cui J,

Gong K, Guo N, Kim K, Liu H, Li Q. Med Image Anal. 2022; 80: 102519

[8] Zezhou Cheng, Matheus Gadelha, Subhransu Maji, Daniel Sheldon, A Bayesian Perspective on the Deep Image Prior, Proceedings of the IEEE Conference on Computer Vision and Pattern Recognition, 2019, pp. 5443-51

[9] Maeda Y, Kudomi N, Yamamoto H, Yamamoto Y, Nishiyama Y. Image accuracy and quality test in rate constant depending on reconstruction algorithms with and without incorporating PSF and TOF in PET imaging. Ann Nucl Med. 2015; 29: pp. 561-9

[10] Watson CC, New, Faster, Image-Based Scatter Correction for 30 PET. IEEE Trans on Nucl Sci, 2000; 47: 1587-94

[11] Kudomi N, Watabe H, Hayashi T, Oka H, Miyake Y, Iida H. Optimization of transmission scan duration for $^{15}$O PET study with sequential dual tracer administration using N-index. Ann Nucl Med. 2010; 24: pp. 413-20.

[12] Koeppe RA, Holden JE, Ip WR. Performance comparison of parameter estimation techniques for the quantitation of local cerebral blood flow by dynamic positron computed tomography. J Cereb Blood Flow Metab. 1985; 5:224-34.

[13] Phelps ME, Huang SC, Hoffman EJ, Selin C, Sokoloff L, Kuhl DE. Tomographic measurement of local cerebral glucose metabolic rate in humans with (F-18)2-fluoro-2-deoxy-D-glucose: validation of method. Ann Neurol. 1979; 6: 371-388

[14] Koeppe RA, Holthoff VA, Frey KA, Kilbourn MR, Kuhl DE. Compartmental analysis of [11C]flumazenil kinetics for the estimation of ligand transport rate and receptor distribution using positron emission tomography. J Cereb Blood Flow Metab. 1991; 11: 735-744.

[15] Schiepers C, Dahlbom M, Chen W, Cloughesy T, Czernin J, Phelps ME, Huang SC. Kinetics of 3'-deoxy-3'-18F-fluorothymidine during treatment monitoring of recurrent high-grade glioma. J Nucl Med. 2010; 51: 720-727

[16] Milan Decuyper, Jens Maebe, Roel Van Holen, Stefaan Vandenberghe, Artificial intelligence with deep learning in nuclear medicine and radiology. EJNMMI Phys 2021; 8: 81

[17] Xing Y, Qiao W, Wang T, Wang Y, Li C, Lv Y, Xi C, Liao S, Qian Z, Zhao J. Deep learning-assisted PET imaging achieves fast scan/low-dose examination. EJNMMI Phys. 2022; 9:7

[18] Gerald Bonardel , Axel Dupont , Pierre Decazes, Mathieu Queneau , Romain Modzelewski , Jeremy Coulot, Nicolas Le Calvez, Sébastien Hapdey, Clinical and phantom validation of a deep learning based denoising algorithm for F-18-FDG PET images from lower detection counting in comparison with the standard acquisition. EJNMMI Phys. 2022; 9: 36

[19] Kudomi N, Maeda Y, Hatakeyama T, Yamamoto Y, Nishiyama Y. Fully parametric imaging

with reversible tracer [18]F-FLT within a reasonable time. Radiol Phys Technol. 2017; 10: 41-8

Figure Captions:

Fig. 1: Apparatus of the phantom used in the present study.

ROI, region of interest

Fig. 2: Progress of $CV_{PF}$, $CV_{OvsG}$, $CV_C$ and $CV_F$, against epochs for FBP without filtering in the $22^{nd}$ frame (left) and OSEM+TOF +PSF with 3 mm filter in the $2^{nd}$ frame (right). DIP reconstructed images during the procedure for several epochs are shown above the CV curve panels.

CV, coefficient of variances; FBP, filtered back projection; OSEM, ordered-subsets expectation maximization; TOF, time of flight; PSF, point spread function; DIP, deep image prior

Fig. 3: Relationships in CVs in $CV_C$ and $CV_F$. ROIs in reconstructed images and decay rate images between those with and without the DIP denoising procedure. The dashed line in black and gray indicate $y = x$ and $y = \frac{1}{\sqrt{2}} x$, respectively.

CV, coefficient of variances; ROIs, regions of interest

Fig. 4: Decay rate images for the reconstruction algorithms with/without TOF (and PSF) correction(s) and filters, before and after DIP, obtained from 0 to 60 min and 60 to 120 min, respectively.

TOF, time of flight; PSF, point spread function; DIP, deep image prior; ORG, original

Fig. 5: Profile of the cross-sectional distribution of pixel value in the decay rate from the first (first row) and second halves (second row) for the applied types of reconstruction algorithms. The respective colors in the figures represent the size of filter on the reconstructed images indicated at the top. The horizontal dashed lines indicate decay constants of $^{11}$C and $^{18}$F.

FBP; filtered back projection; TOF; time of flight; OSEM, ordered-subsets expectation maximization; W/O, without; DIP, deep image prior

Table 1: Epochs for generated images were chosen as optimal.

| Frame # | FBP | | | FBPTOF | | | OSEM | | | OSEM_TOFPSF | | |
|---|---|---|---|---|---|---|---|---|---|---|---|---|
| | W/O | 1 mm | 3 mm | W/O | 1 mm | 3 mm | W/O | 1 mm | 3 mm | W/O | 1 mm | 3 mm |
| 1 | 510 | 2180 | 7830 | 480 | 1420 | 4990 | 860 | 1130 | 1440 | 1120 | 2600 | 6260 |
| 2 | 400 | 470 | 4560 | 640 | 3160 | 7030 | 2810 | 840 | 3390 | 900 | 9800 | 9490 |
| 3 | 1780 | 350 | 4620 | 520 | 990 | 6680 | 790 | 860 | 4260 | 760 | 1130 | 6360 |
| 4 | 320 | 870 | 2560 | 540 | 760 | 2650 | 530 | 570 | 3370 | 680 | 1710 | 9310 |
| 5 | 480 | 490 | 1750 | 400 | 1020 | 1560 | 460 | 760 | 2160 | 470 | 780 | 8420 |
| 6 | 600 | 390 | 1550 | 480 | 1050 | 2340 | 360 | 740 | 9640 | 700 | 520 | 2090 |
| 7 | 550 | 420 | 4700 | 390 | 450 | 9140 | 470 | 720 | 5570 | 470 | 600 | 5260 |
| 8 | 240 | 310 | 1740 | 530 | 1220 | 1110 | 500 | 510 | 9380 | 310 | 830 | 8800 |
| 9 | 230 | 350 | 1320 | 480 | 800 | 5220 | 450 | 490 | 1830 | 600 | 4610 | 4690 |
| 10 | 790 | 380 | 5670 | 720 | 510 | 9260 | 400 | 740 | 8990 | 960 | 600 | 7480 |
| 11 | 930 | 1120 | 970 | 400 | 800 | 4310 | 270 | 560 | 5450 | 300 | 400 | 3410 |
| 12 | 720 | 310 | 1170 | 580 | 1190 | 3360 | 2370 | 710 | 4090 | 380 | 1060 | 4950 |
| 13 | 680 | 190 | 4240 | 370 | 700 | 1260 | 250 | 900 | 3320 | 320 | 520 | 6140 |
| 14 | 820 | 320 | 1320 | 370 | 550 | 7160 | 430 | 650 | 790 | 600 | 560 | 3460 |
| 15 | 780 | 490 | 7670 | 240 | 670 | 1020 | 380 | 630 | 4290 | 480 | 580 | 3240 |
| 16 | 180 | 290 | 580 | 440 | 460 | 6200 | 480 | 460 | 2830 | 520 | 720 | 1900 |
| 17 | 1050 | 300 | 680 | 1250 | 540 | 2110 | 310 | 500 | 8830 | 530 | 480 | 2800 |
| 18 | 1240 | 290 | 520 | 230 | 500 | 8140 | 210 | 480 | 1520 | 390 | 600 | 7320 |
| 19 | 1630 | 250 | 4430 | 290 | 500 | 4780 | 300 | 510 | 4490 | 370 | 880 | 2180 |
| 20 | 450 | 450 | 8240 | 300 | 310 | 1710 | 450 | 340 | 7570 | 320 | 700 | 2720 |
| 21 | 750 | 390 | 710 | 360 | 330 | 7850 | 190 | 430 | 1820 | 270 | 410 | 6510 |
| 22 | 1150 | 220 | 4890 | 610 | 450 | 3810 | 320 | 660 | 8260 | 250 | 320 | 1260 |
| 23 | 2190 | 540 | 680 | 270 | 220 | 3210 | 270 | 600 | 7500 | 280 | 450 | 7320 |
| 24 | 2150 | 310 | 2470 | 160 | 1880 | 840 | 260 | 250 | 810 | 220 | 600 | 5400 |
| Mean | 859 | 486 | 3119 | 460 | 853 | 4405 | 588 | 626 | 4650 | 508 | 1310 | 5282 |

W/O: Without filter, 1 mm and 1 mm: Gaussian filter sized with 3mm and 5 mm, respectively. FBP, filtered back projection; TOF, time of flight; OSEM, ordered-subsets expectation maximization; PSF, point spread function

Table 2 Mean ± SD ($10^{-3}$ min$^{-1}$) of decay rate for ROIs filled with $^{18}$F ($\lambda = 6.31 \times 10^{-3}$ min$^{-1}$) and $^{11}$C ($\lambda = 34.1 \times 10^{-3}$ min$^{-1}$) regions, respectively

| | FBP | | | FBP+TOF | | | OSEM | | | OSEM+TOF+PSF | | |
|---|---|---|---|---|---|---|---|---|---|---|---|---|
| | WO | 1mm | 3mm | WO | 1mm | 3mm | WO | 1mm | 3mm | WO | 1mm | 3mm |
| $^{18}$F | | | | | | | | | | | | |
| Original | | | | | | | | | | | | |
| 1st | 7.72±4.38 | 7.55±2.57 | 7.18±0.52 | 7.02±2.85 | 7.02±1.55 | 6.83±0.35 | 7.40±3.07 | 7.34±1.67 | 7.06±0.36 | 7.51±2.42 | 7.49±1.29 | 7.28±0.28 |
| 2nd | 6.60±4.11 | 6.64±2.61 | 6.85±0.57 | 6.21±3.91 | 6.28±2.78 | 6.55±0.57 | 6.69±3.21 | 6.81±2.11 | 6.80±0.61 | 5.99±3.10 | 6.00±2.08 | 5.94±0.52 |
| DIP | | | | | | | | | | | | |
| 1st | 7.12±1.87 | 6.78±0.55 | 6.46±0.39 | 6.44±0.72 | 7.22±0.64 | 6.36±0.26 | 6.65±0.90 | 6.94±0.80 | 7.13±0.19 | 7.53±0.68 | 7.77±0.54 | 6.99±0.26 |
| 2nd | 7.03±2.37 | 7.10±0.56 | 7.21±0.24 | 6.29±0.76 | 6.94±0.51 | 6.74±0.16 | 6.57±0.85 | 7.04±0.80 | 6.57±0.32 | 6.19±0.51 | 5.53±0.81 | 5.90±0.06 |
| $^{11}$C | | | | | | | | | | | | |
| Original | | | | | | | | | | | | |
| 1st | 33.87±8.01 | 33.52±3.81 | 33.50±0.66 | 33.64±6.23 | 33.39±2.99 | 33.51±0.44 | 33.97±4.15 | 33.87±2.39 | 33.92±0.52 | 33.50±4.46 | 33.36±2.55 | 33.50±0.61 |
| 2nd | 25.93±30.92 | 31.81±29.72 | 32.20±4.82 | 26.30±27.11 | 29.05±13.61 | 31.60±4.05 | 29.65±14.56 | 28.56±6.21 | 27.56±1.92 | 29.64±12.24 | 28.69±6.10 | 28.21±2.08 |
| DIP | | | | | | | | | | | | |
| 1st | 34.17±2.64 | 35.16±1.54 | 32.82±0.53 | 33.53±1.73 | 33.53±1.06 | 33.08±0.55 | 34.02±1.70 | 33.66±1.04 | 33.90±0.26 | 33.17±1.73 | 33.48±1.23 | 33.51±0.35 |
| 2nd | 43.38±37.04 | 37.29±7.84 | 36.10±4.83 | 34.98±6.31 | 35.15±8.80 | 33.44±3.26 | 30.17±4.06 | 30.89±4.26 | 26.17±2.26 | 32.61±4.00 | 30.75±2.82 | 28.47±2.20 |

W/O: Without filter, 1 mm and 3 mm: Gaussian filter with 1 mm and 3 mm, respectively. FBP, filtered back projection; TOF, time of flight; OSEM, ordered-subsets expectation maximization; PSF, point spread function; DIP, deep image prior; SD, standard deviation.

# Fig 1

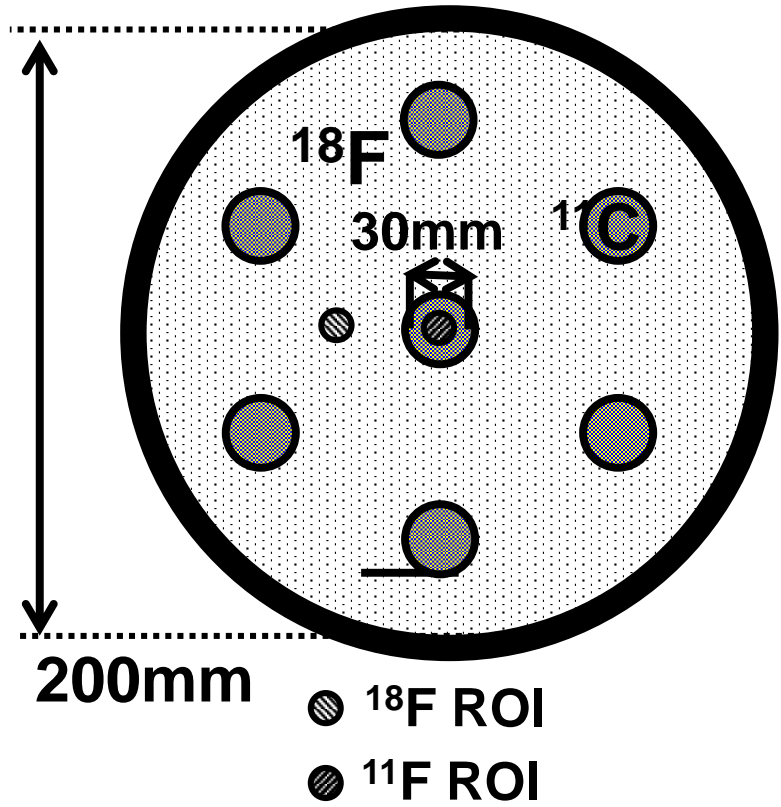

200mm

30mm

$^{18}$F

$^{11}$C

⬤ $^{18}$F ROI

⬤ $^{11}$F ROI

# Fig 2

## FBP without filter 22$^{nd}$ frame

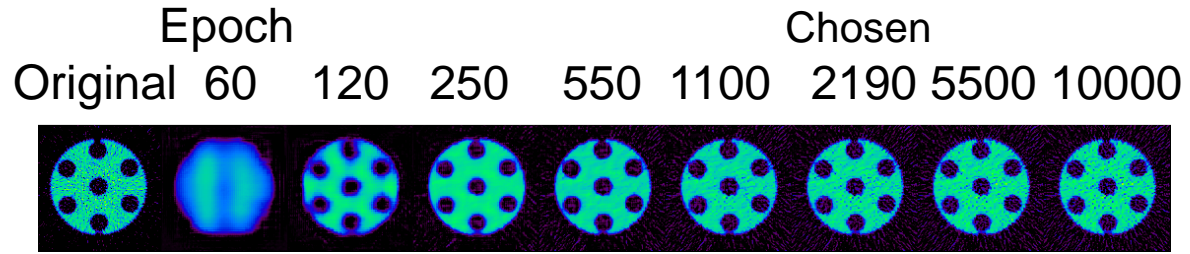

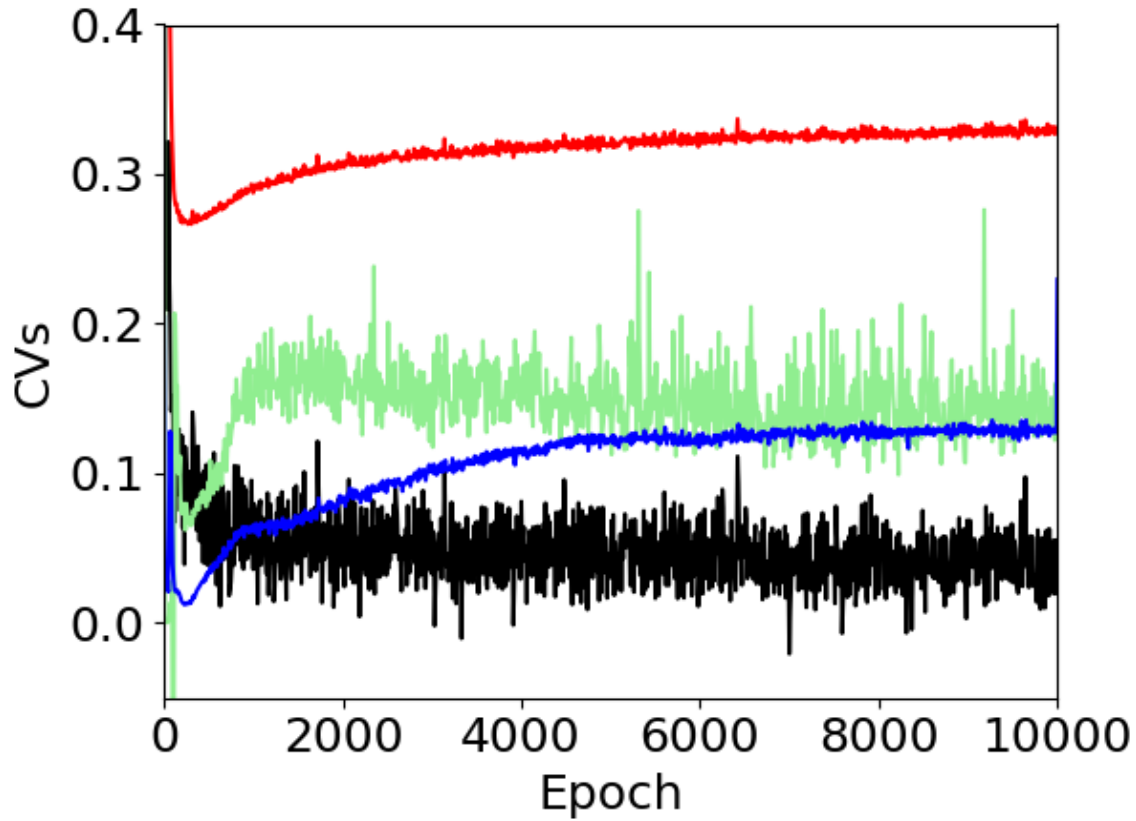

## OSEM+TOF+PSF with 3 mm filter 22$^{nd}$ frame

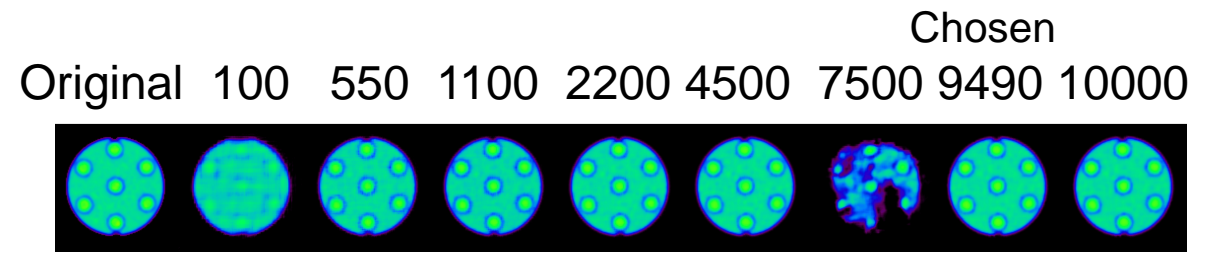

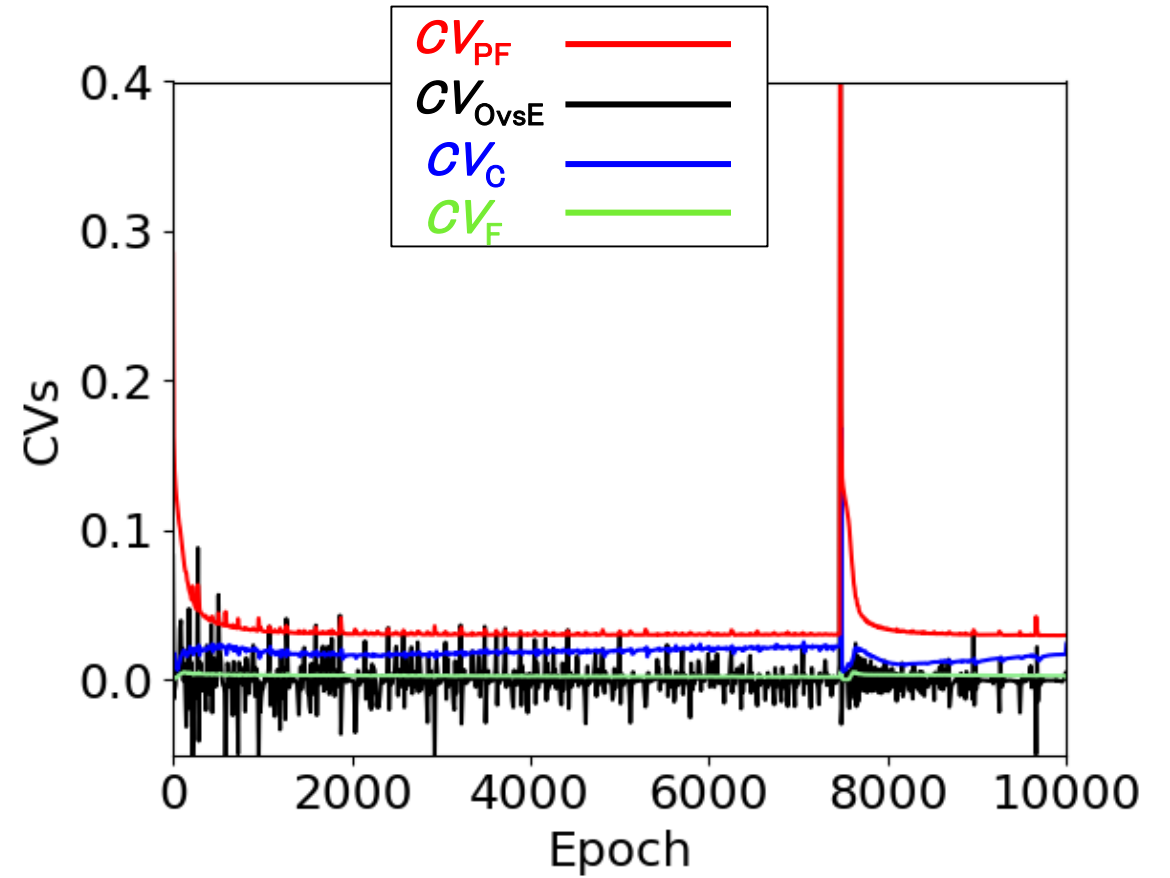

Fig 3

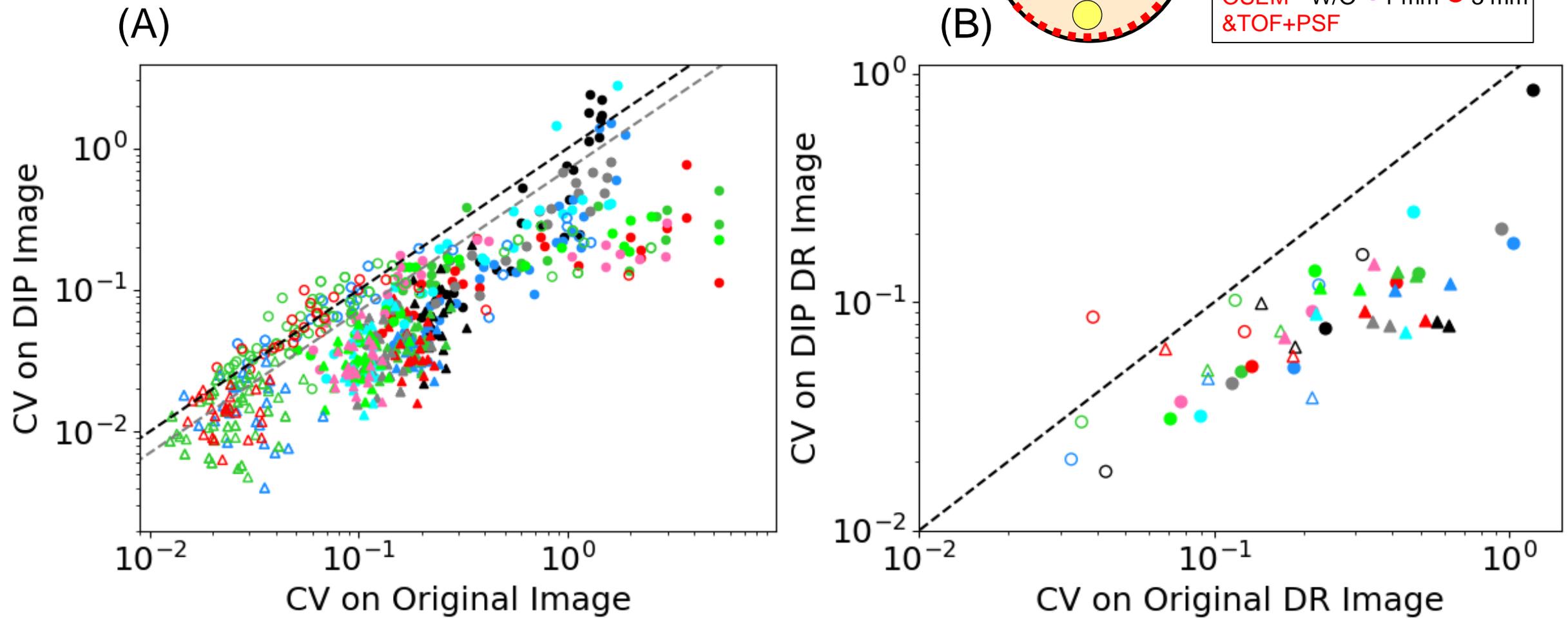

Fig 4

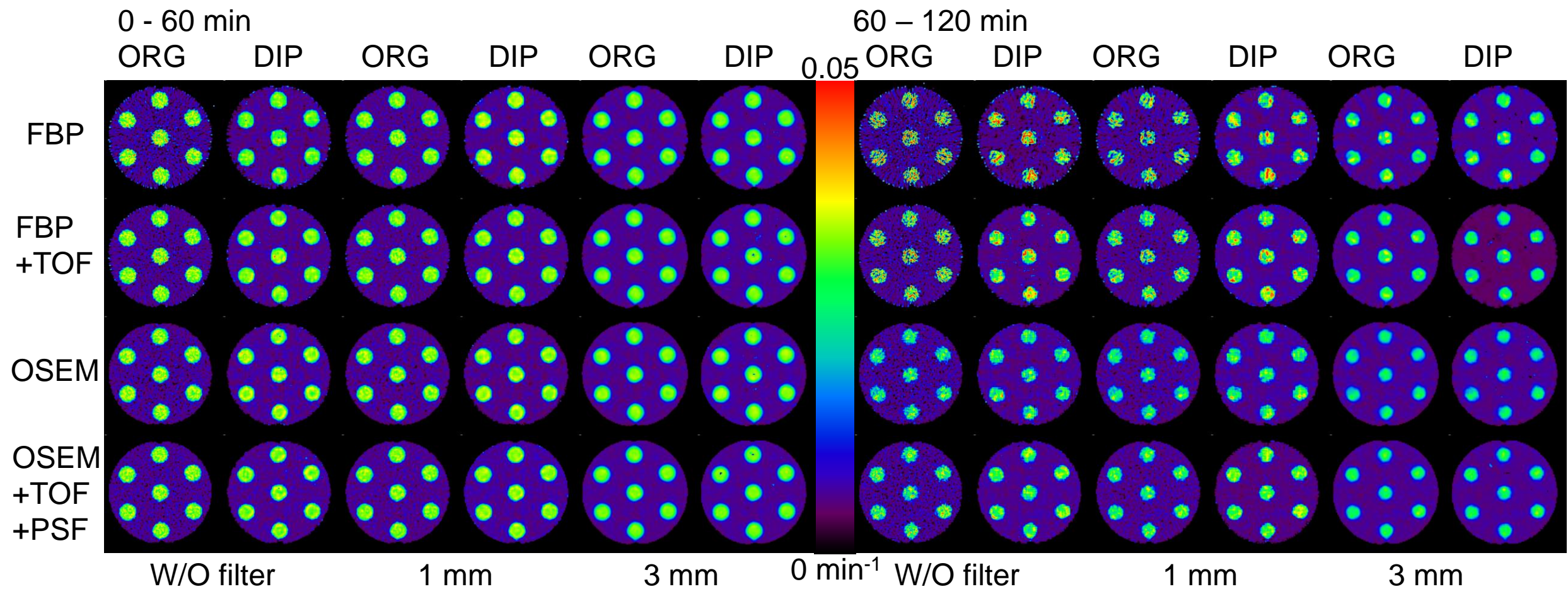

Fig 5

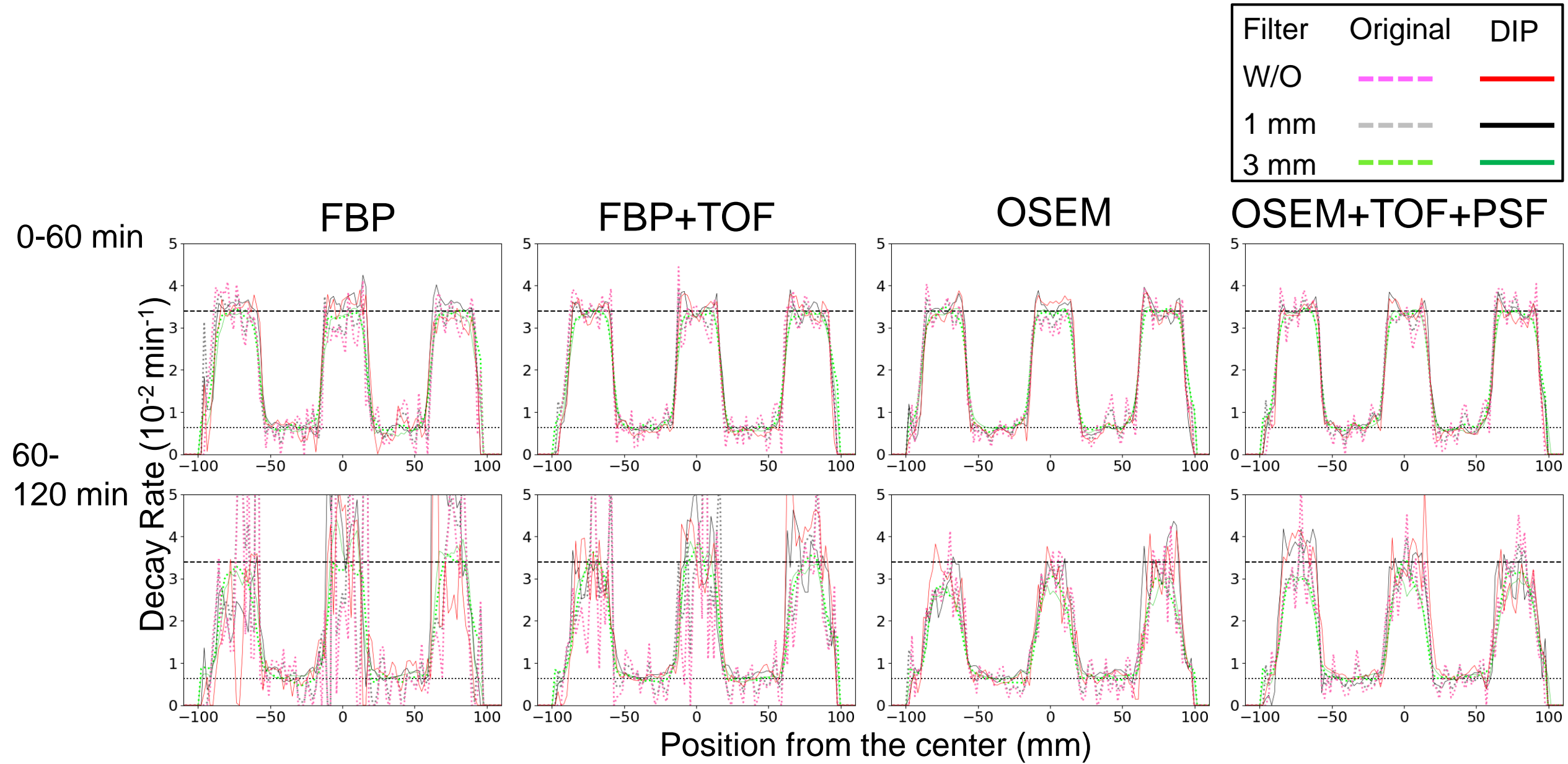

Figure Captions for supplemental material:

Fig S1: Progress of $CV_{PF}$, $CV_{OvsG}$, $CV_C$ and $CV_F$, against epoch for FBP and OSEM reconstruction with/without filtering and TOF and PSF. The numbers indicated are frame numbers.

Fig. S2: DIP generating images. The left is original, middle is chosen as optimal and right is the last (epoch = 10000) images, respectively. The others are half (left to chosen) and twice of epochs (right to chosen) against the chosen ones, respectively, except the epoch was 900 when the epoch for chosen was more than 500. The numbers indicated are frame numbers.

# FigS1(a): FBP without filter

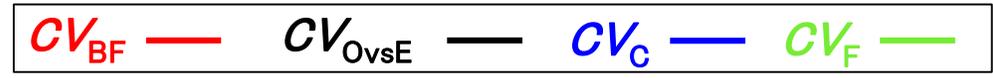

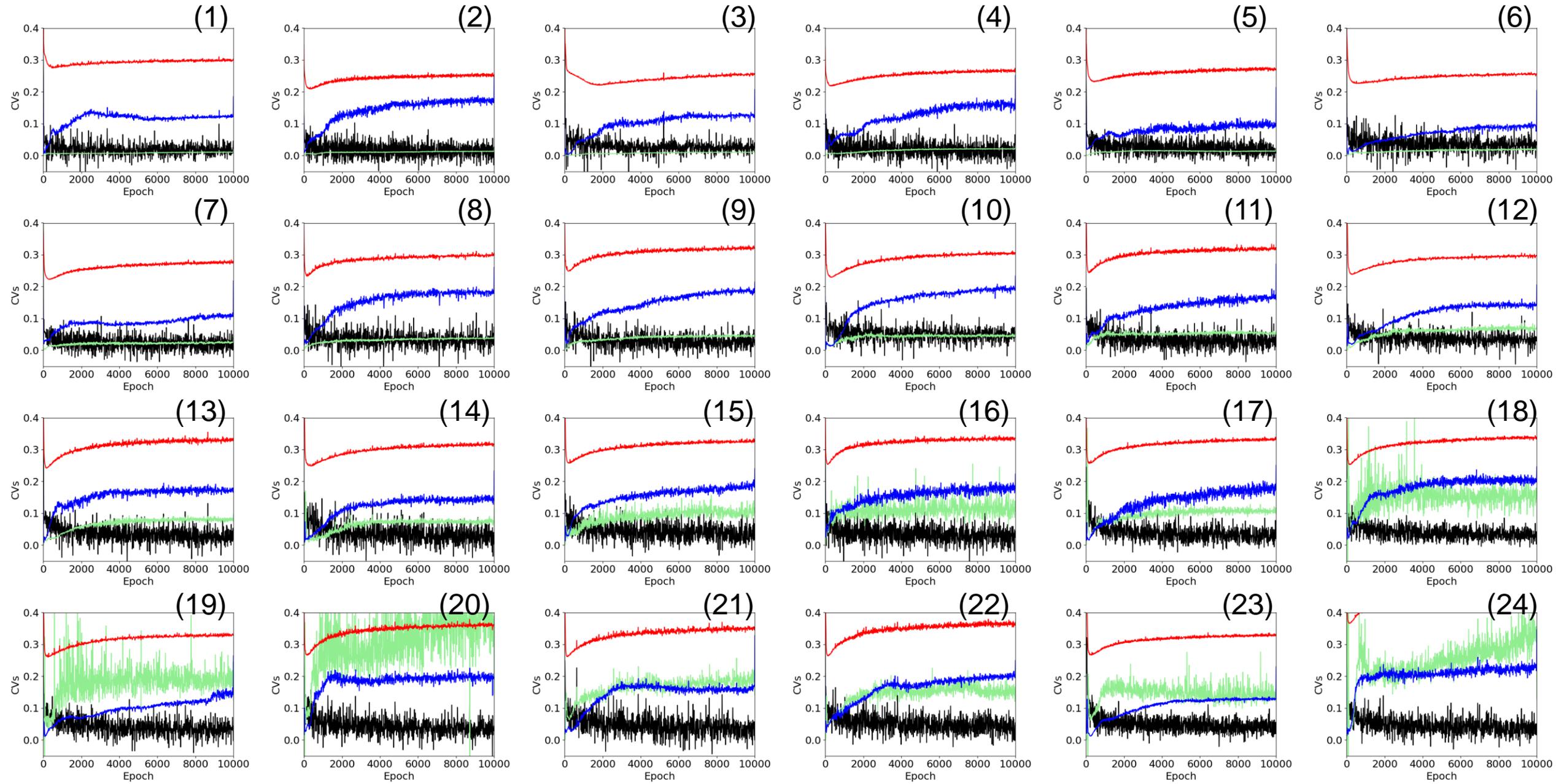

# FigS1(b): FBP with 1 mm filter

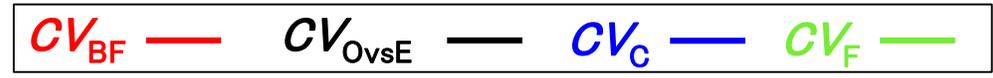

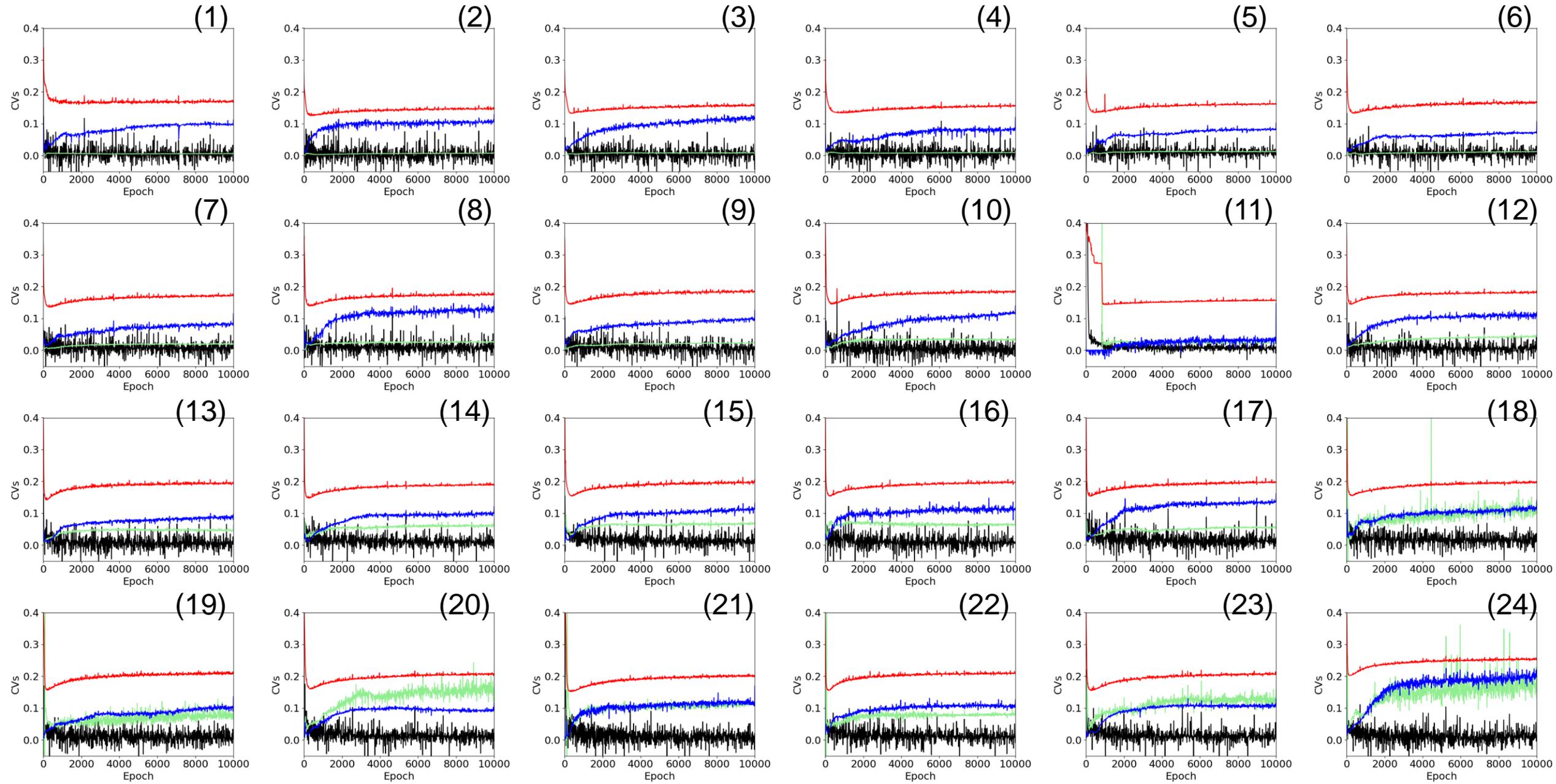

# FigS1(c): FBP with 3 mm filter

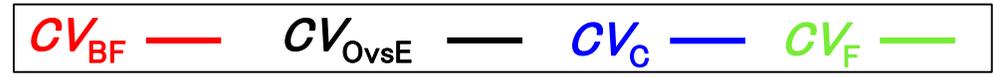

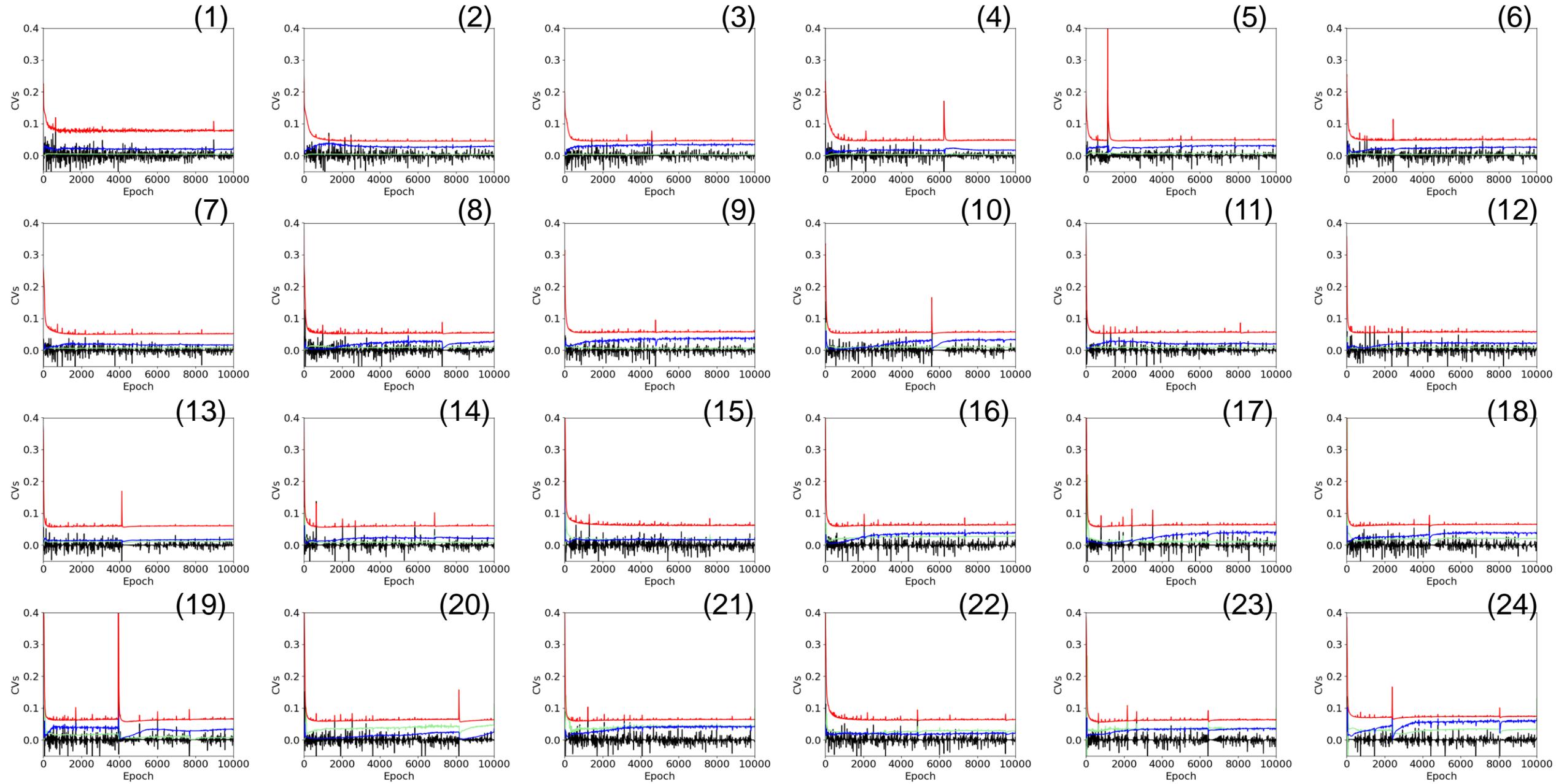

# FigS1(d): FBP+TOF without filter

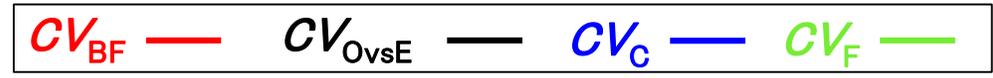

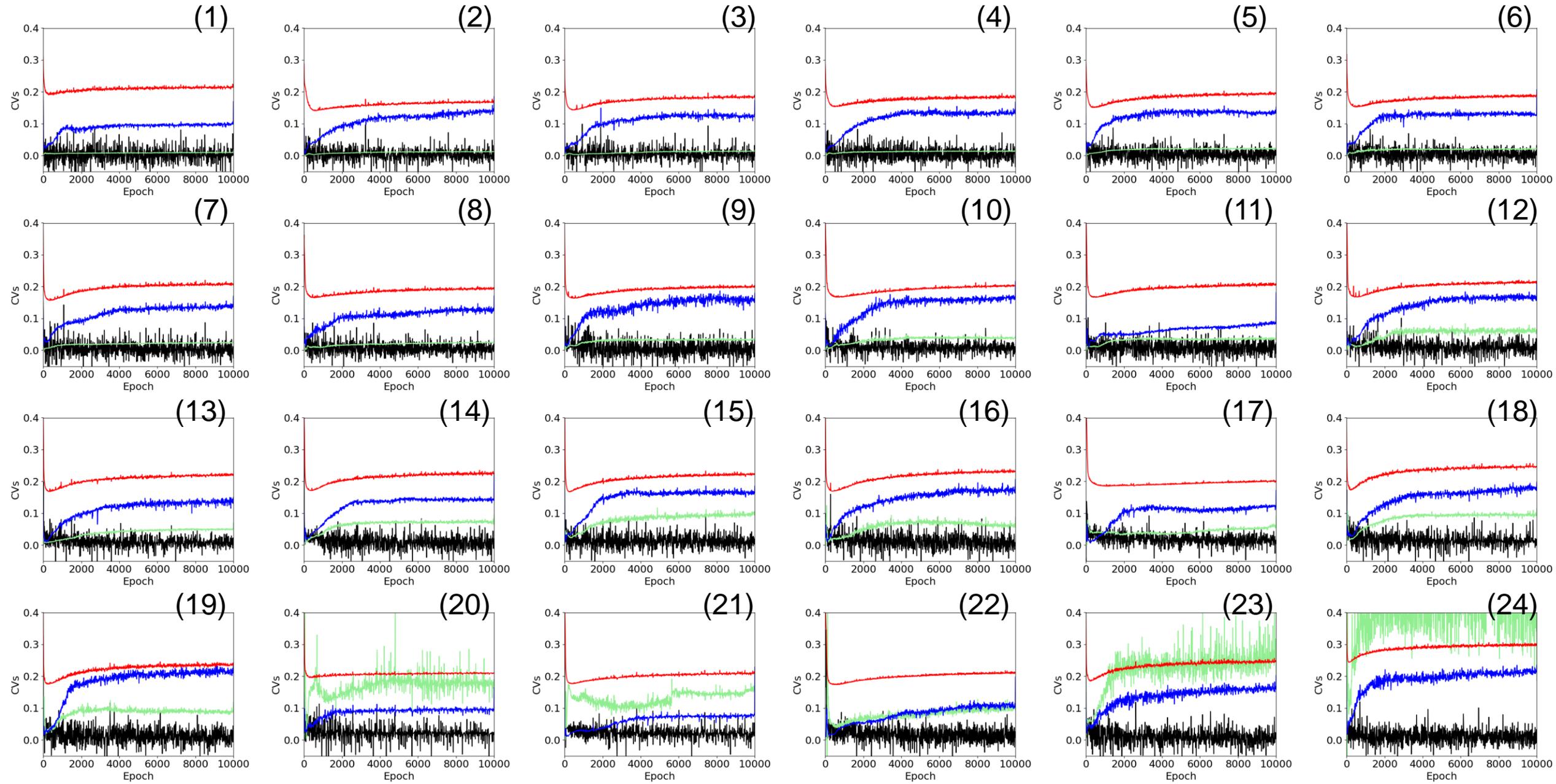

# FigS1(e): FBP+TOF with 1 mm filter

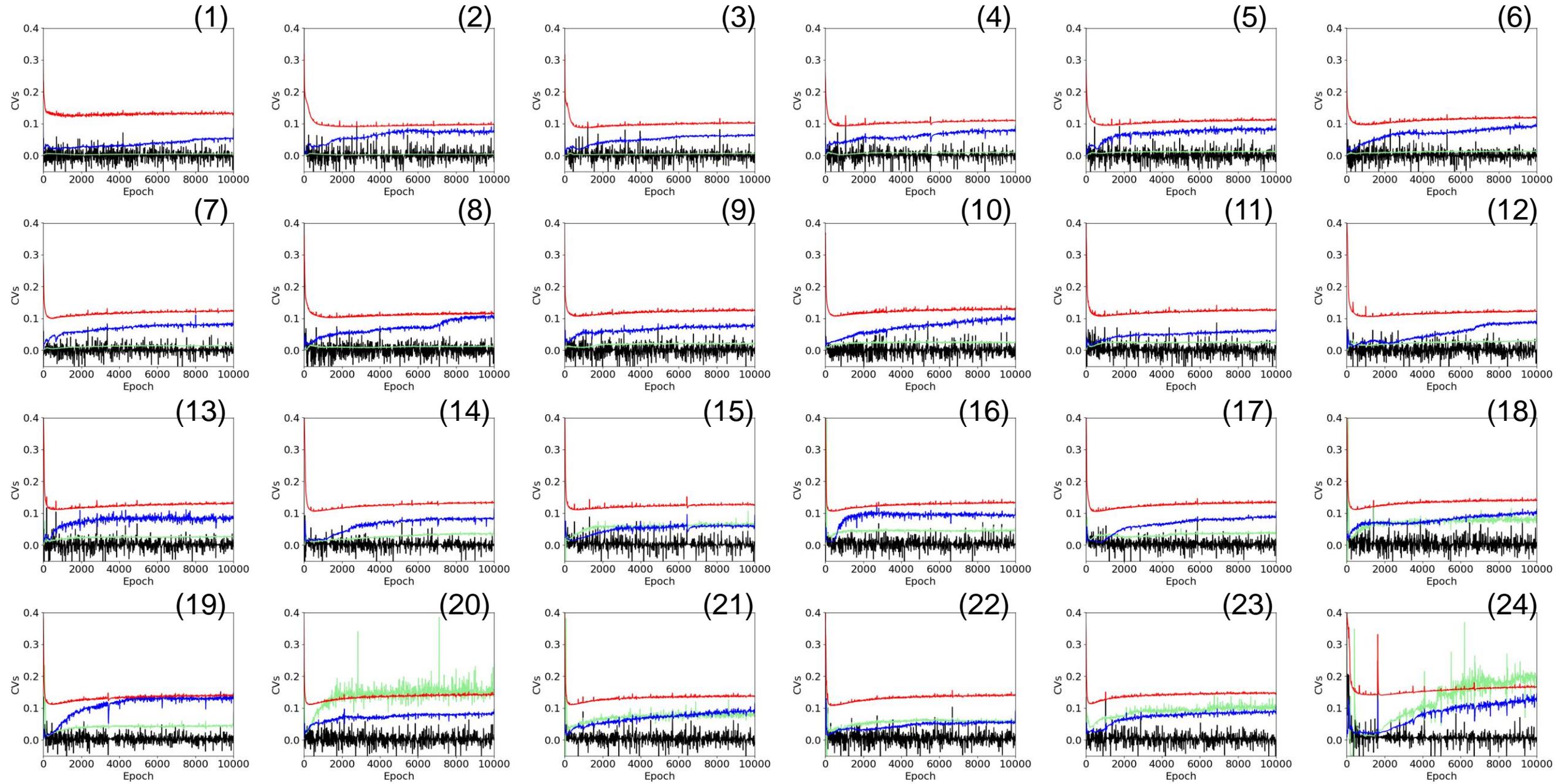

# FigS1(f): FBP+TOF with 3 mm filter

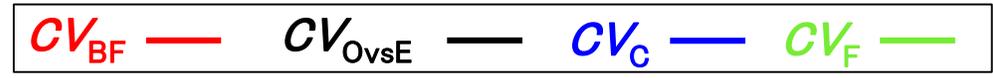

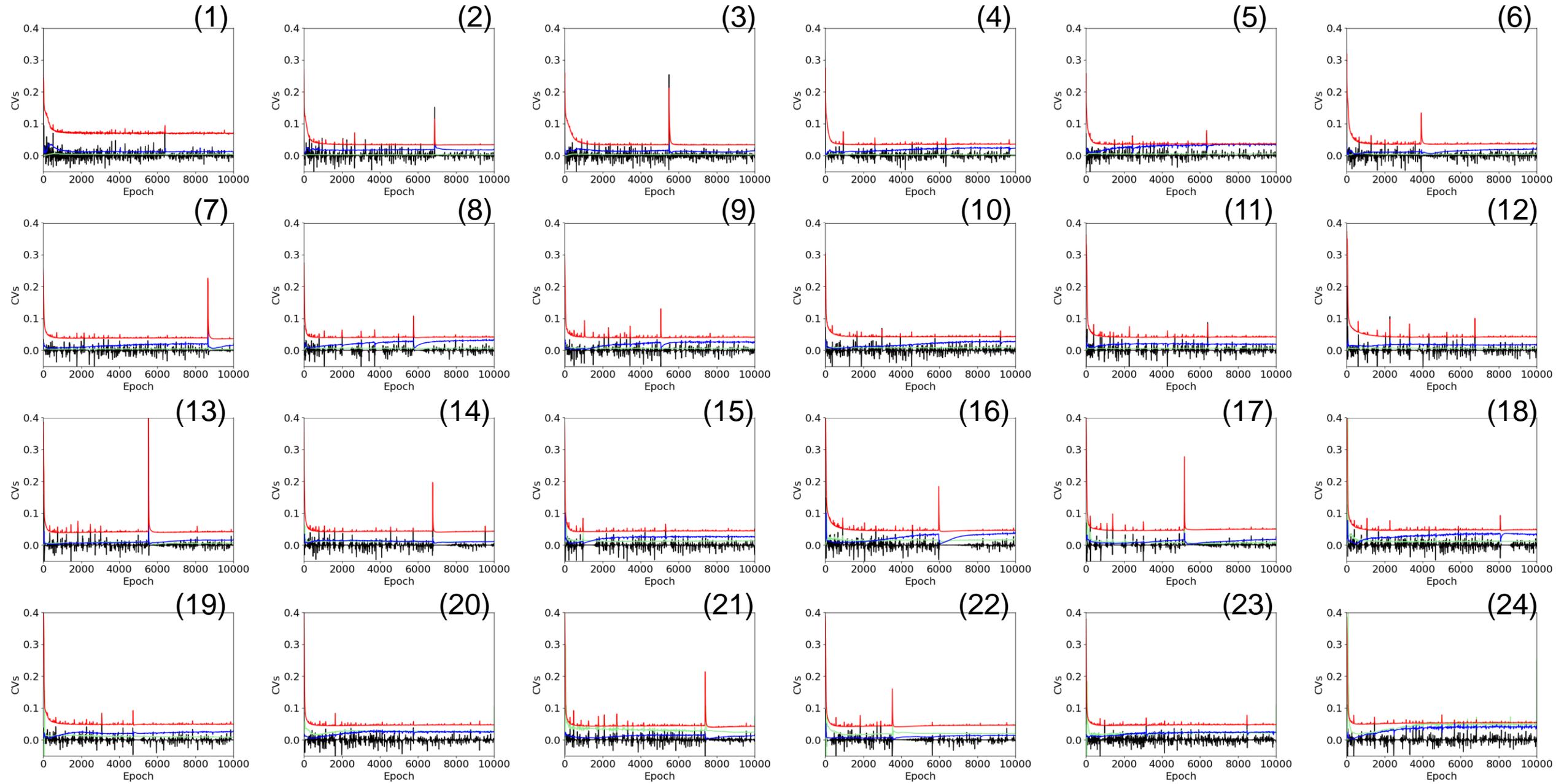

# FigS1(g): OSEM without filter

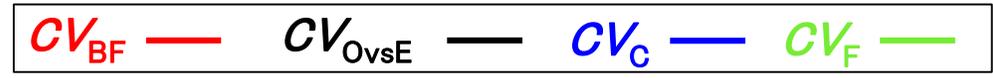

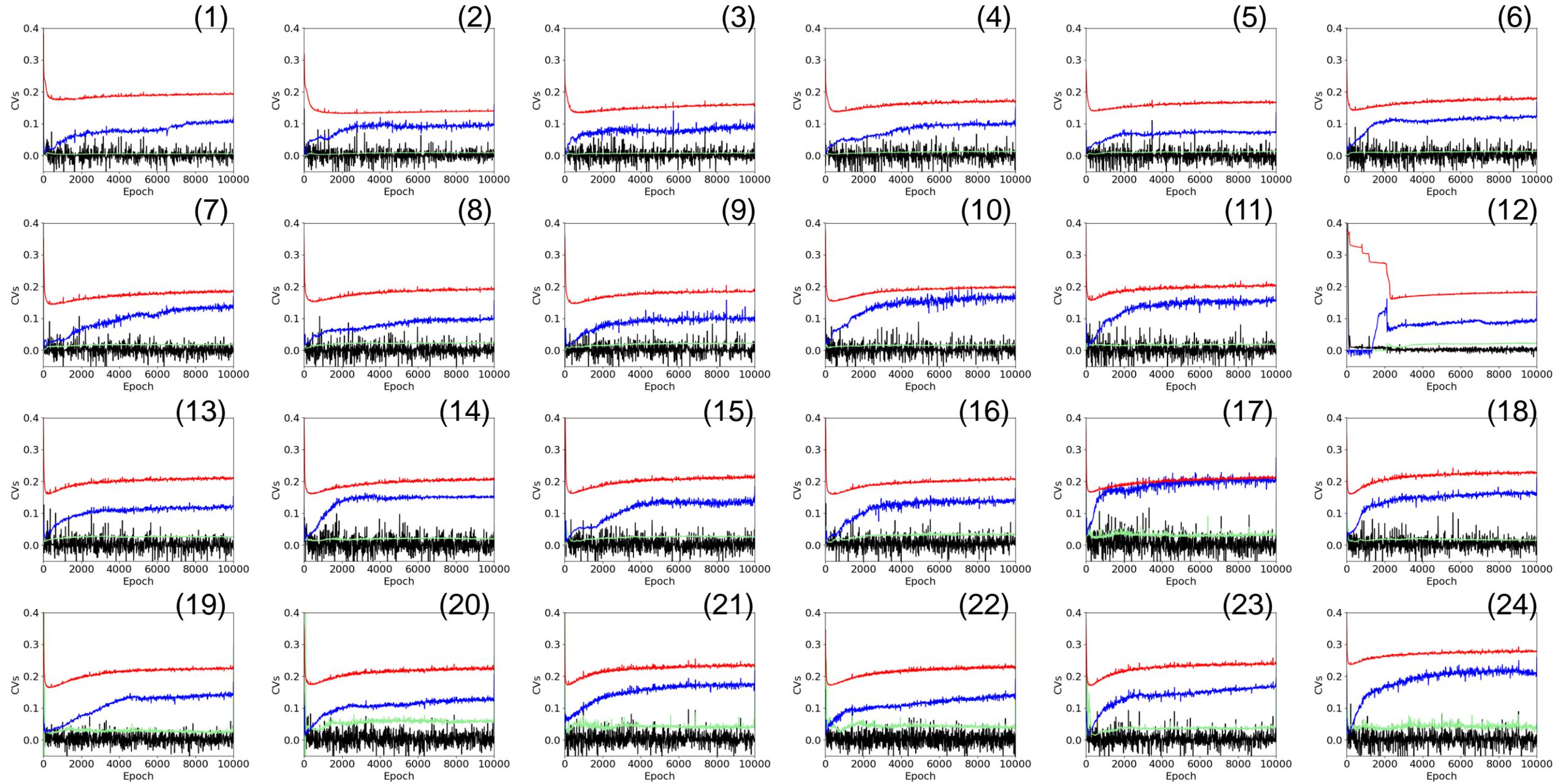

# FigS1(h): OSEM with 1 mm filter

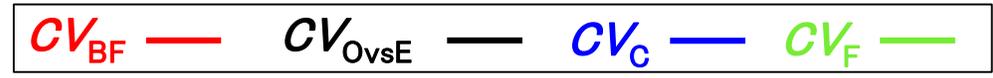

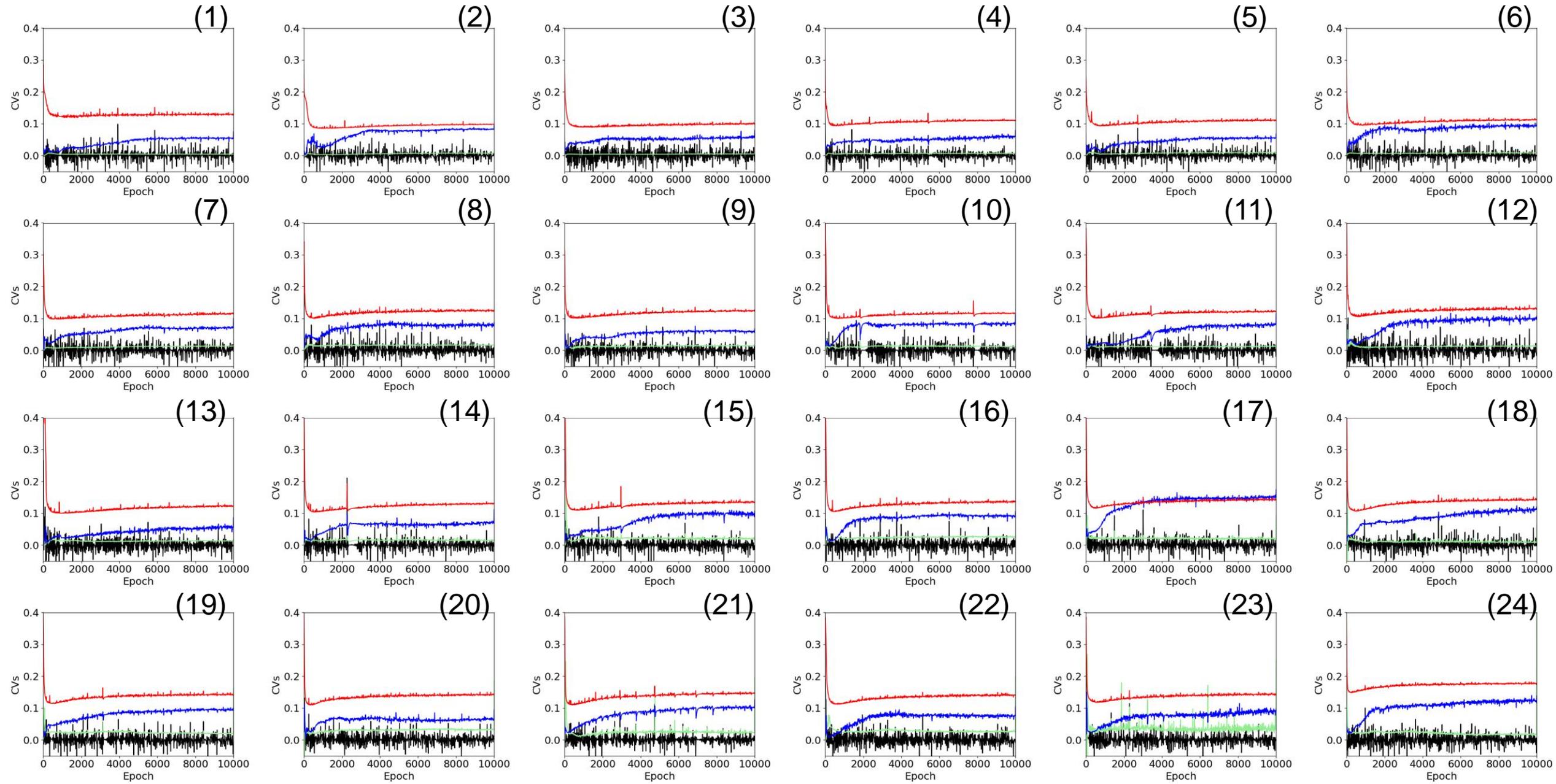

# FigS1(i): OSEM with 3 mm filter

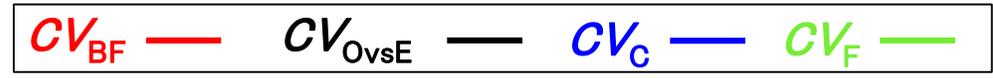

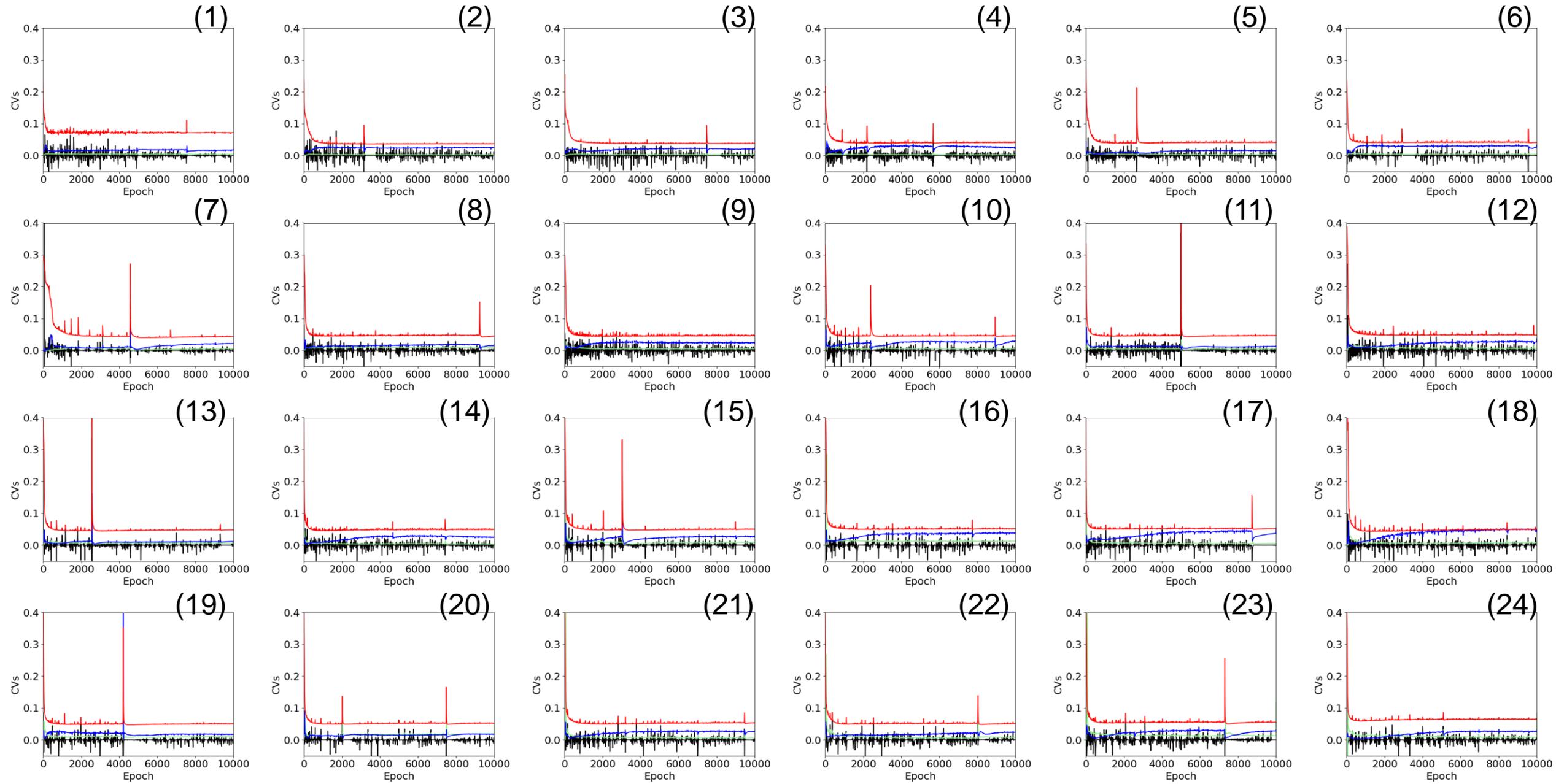

# FigS1(j): OSEM+TOF+PSF without filter

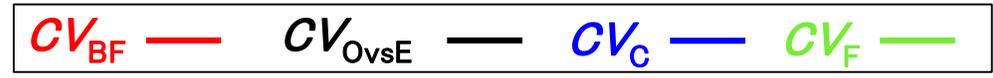

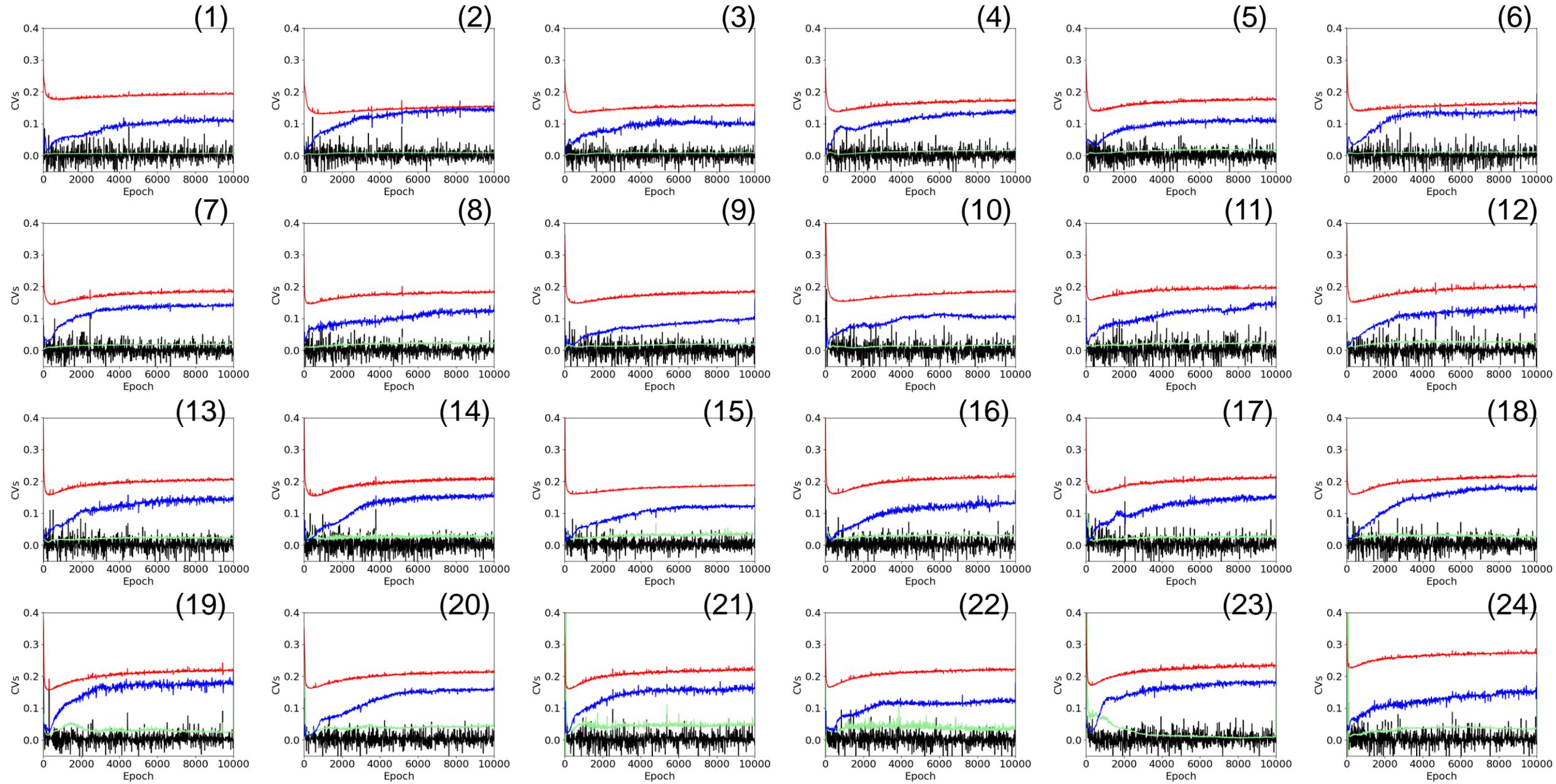

# FigS1(k): OSEM+TOF+PSF with 1 mm filter

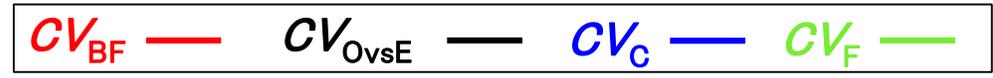

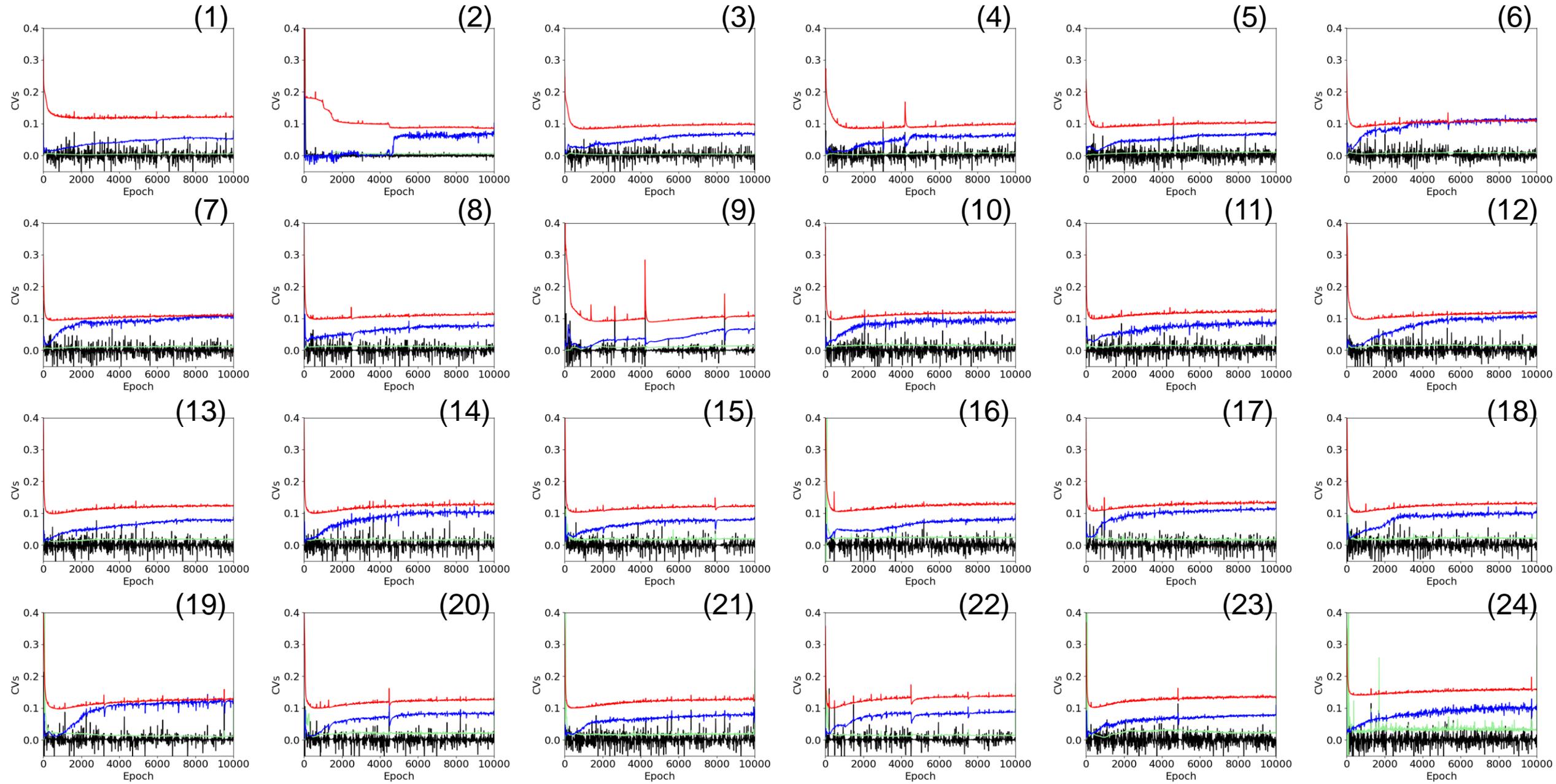

# FigS1(l): OSEM+TOF+PSF with 3 mm filter

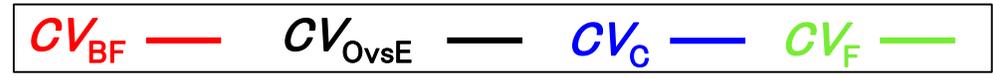

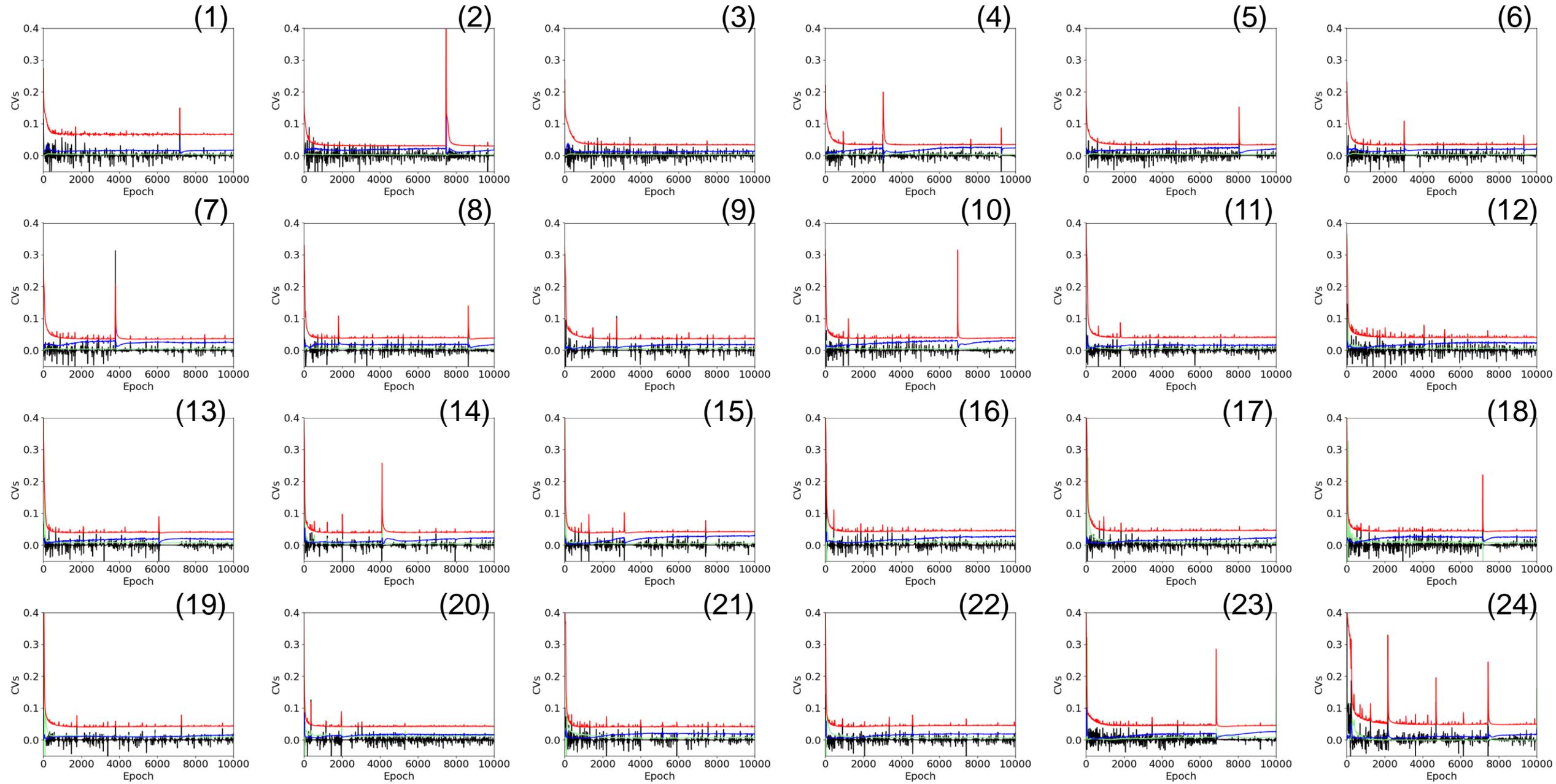

FigS2(a): FBP without filter

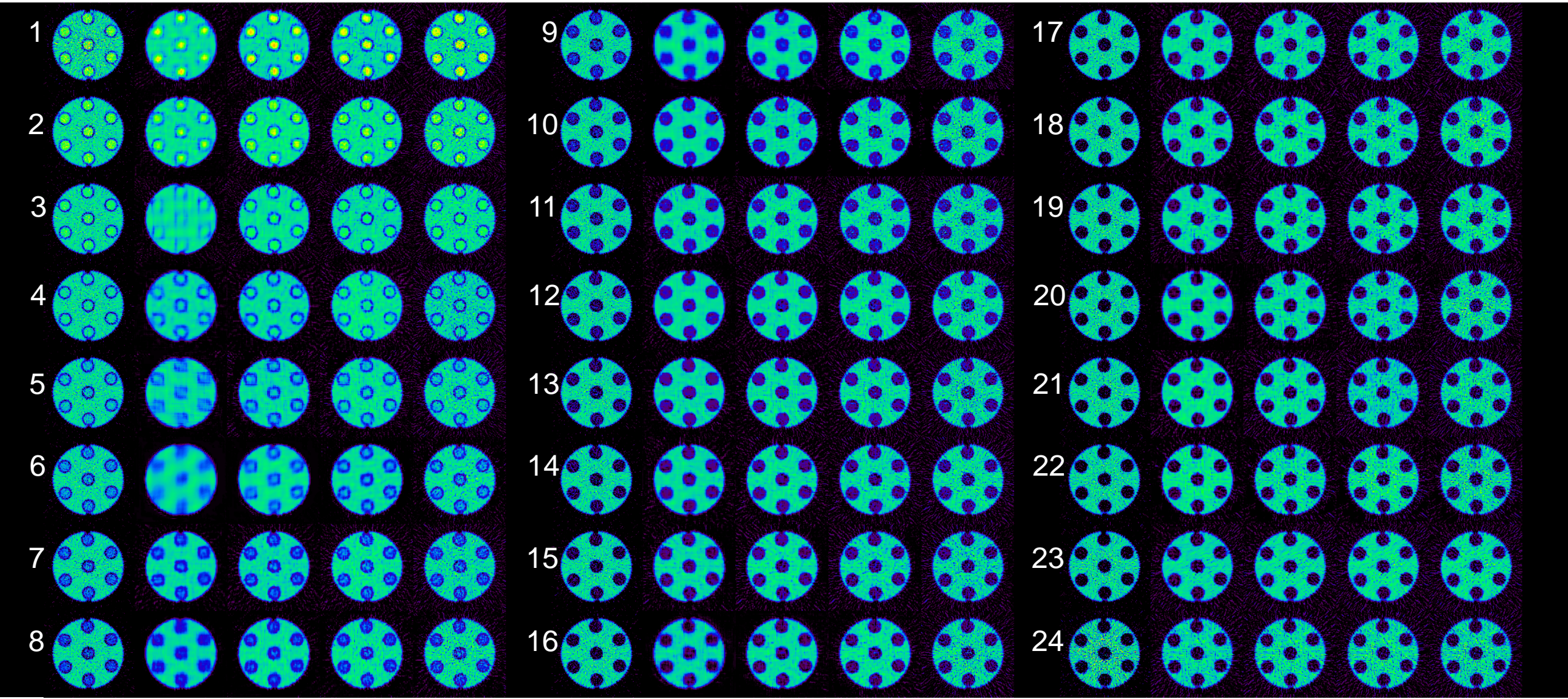

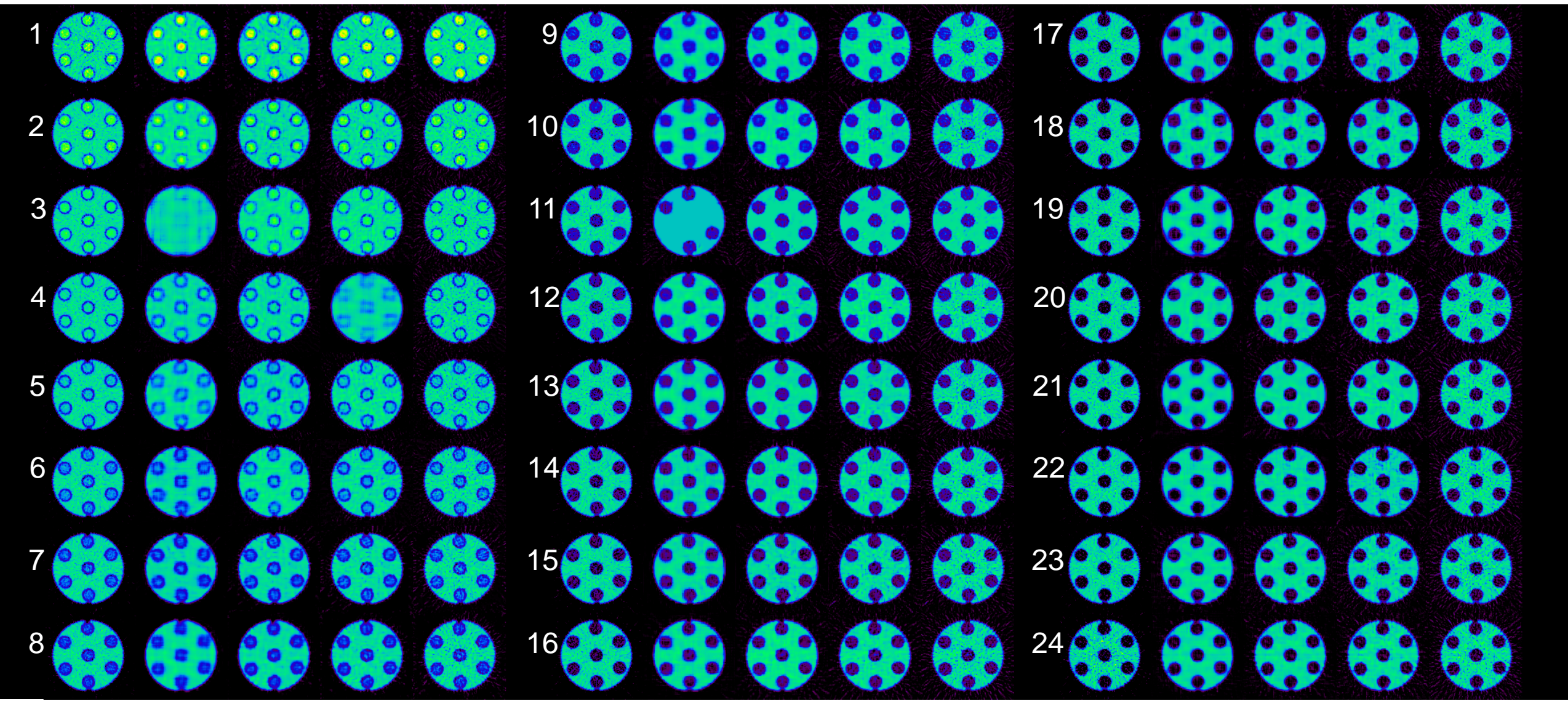

FigS2(a): FBP with 3 mm filter

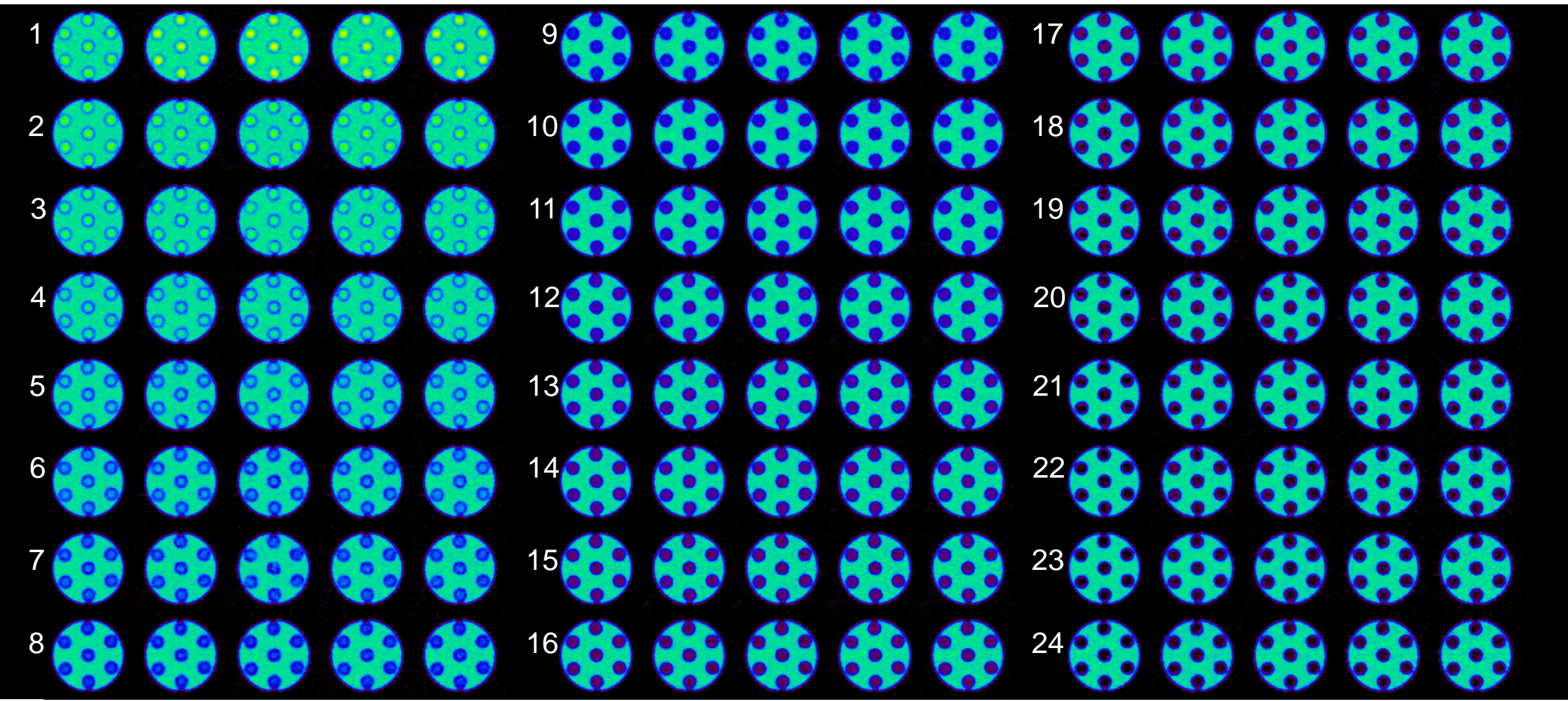

# FigS2(a): FBP+TOF without filter

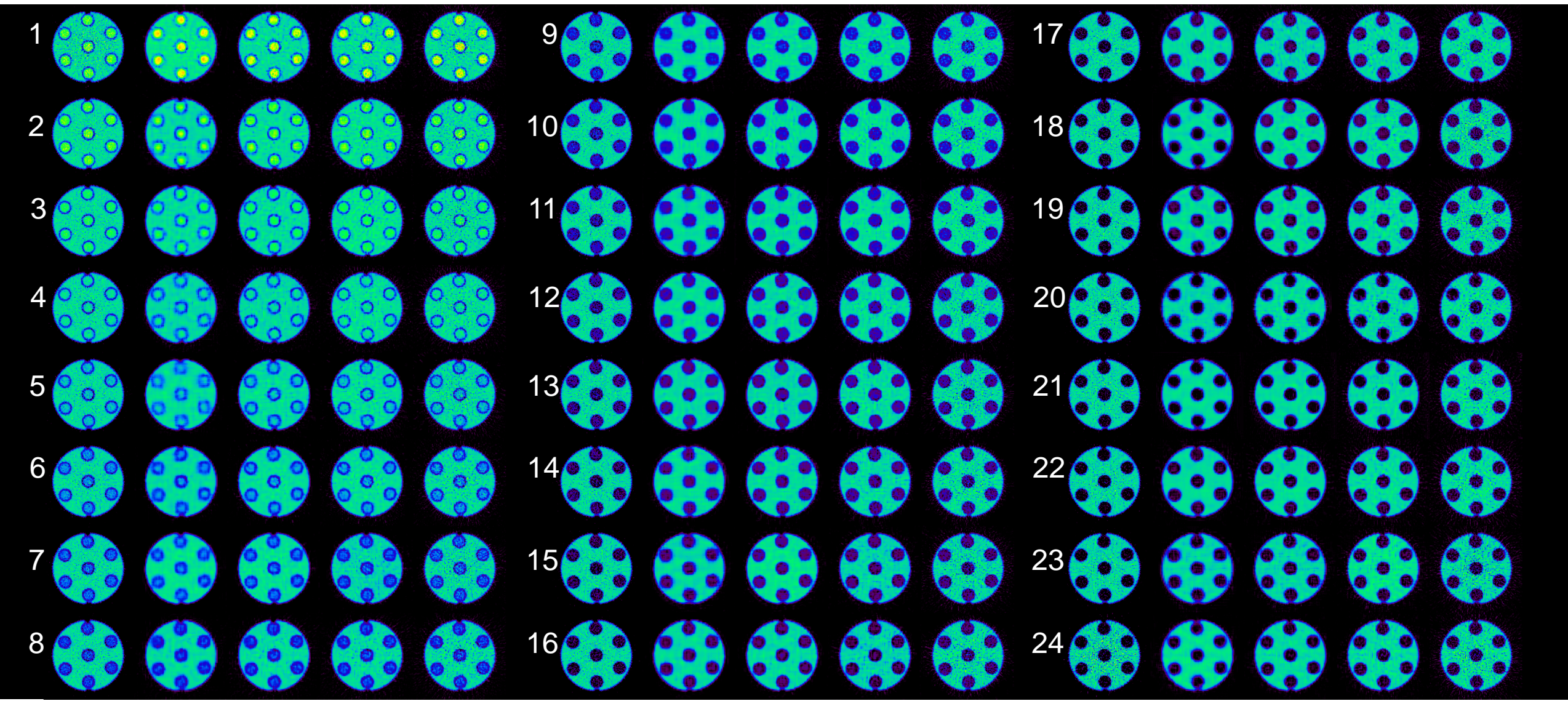

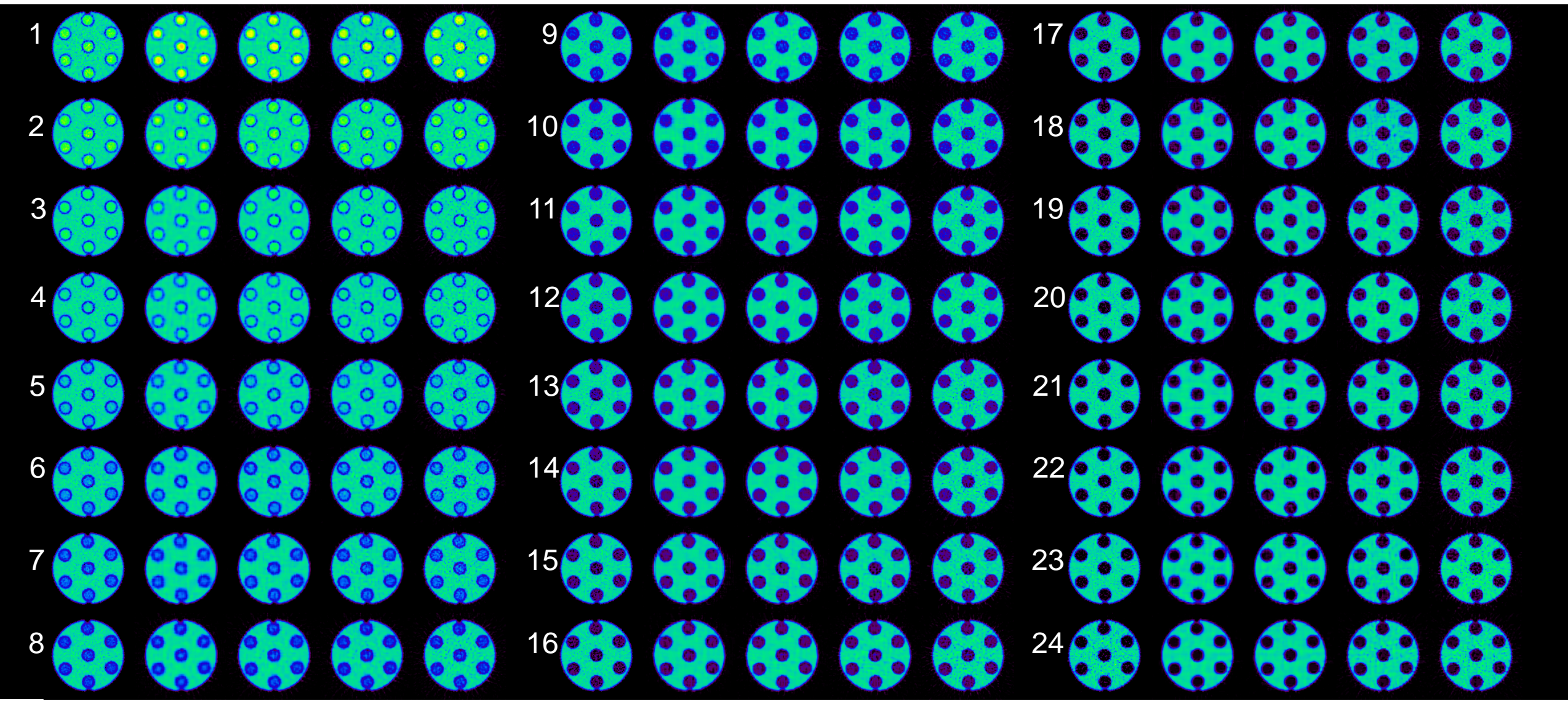

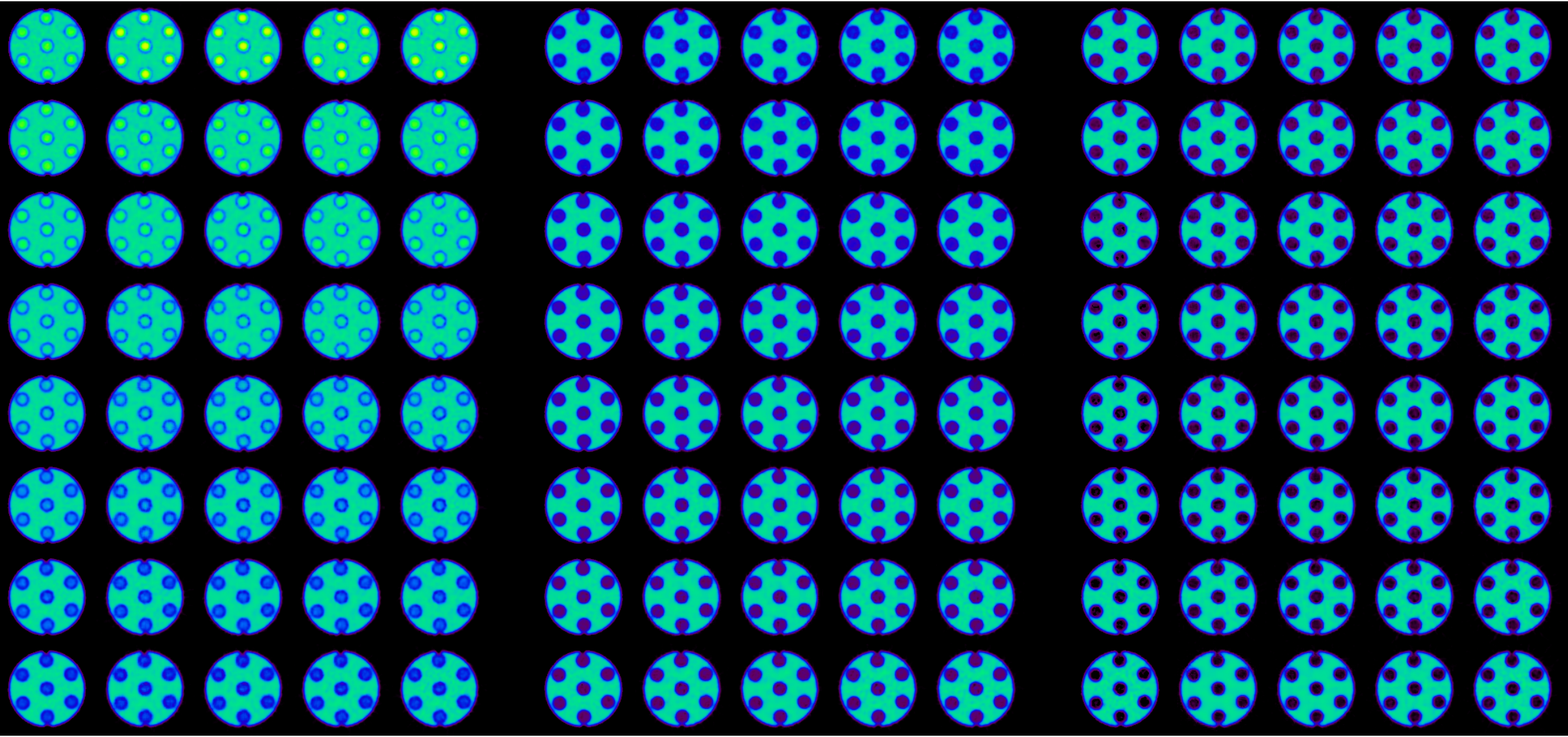

# FigS2(a): OSEM without filter

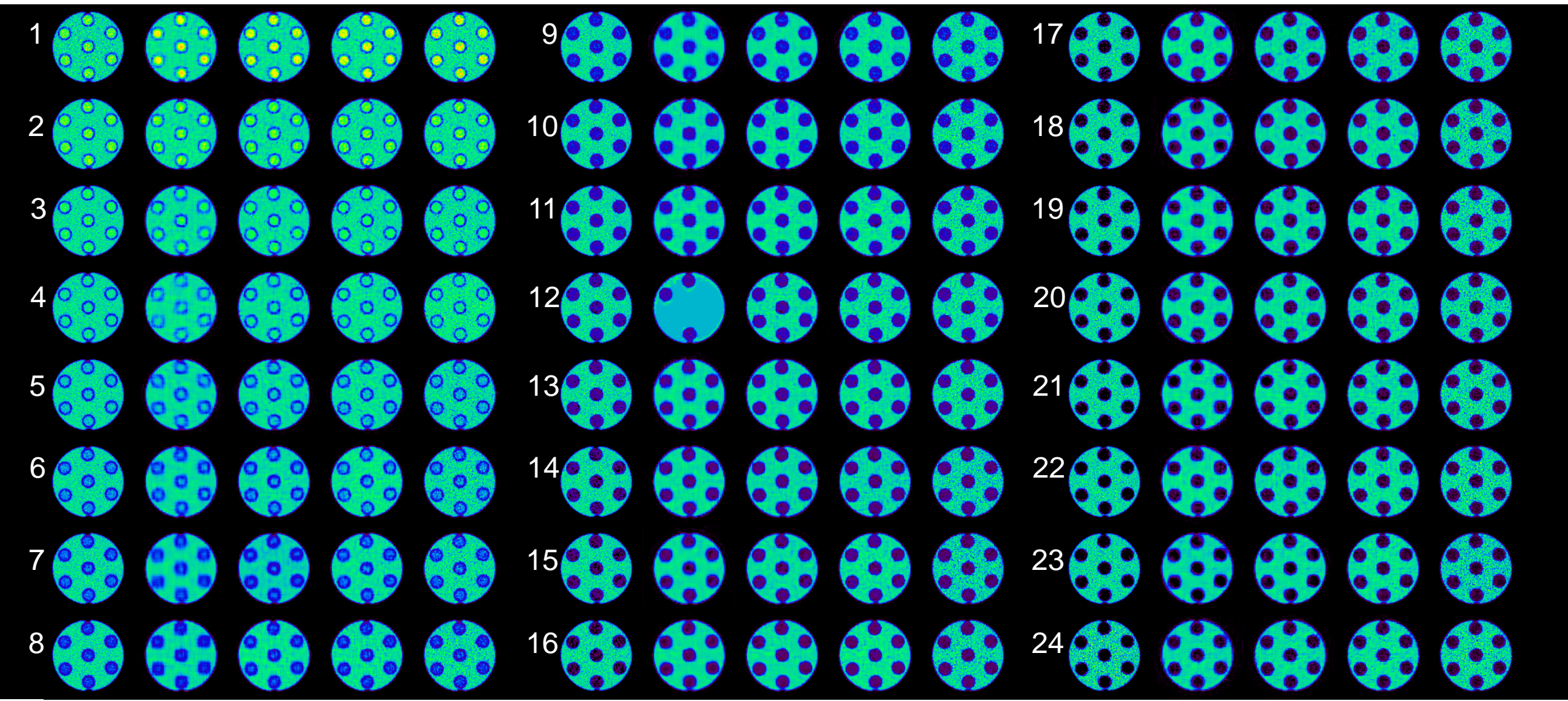

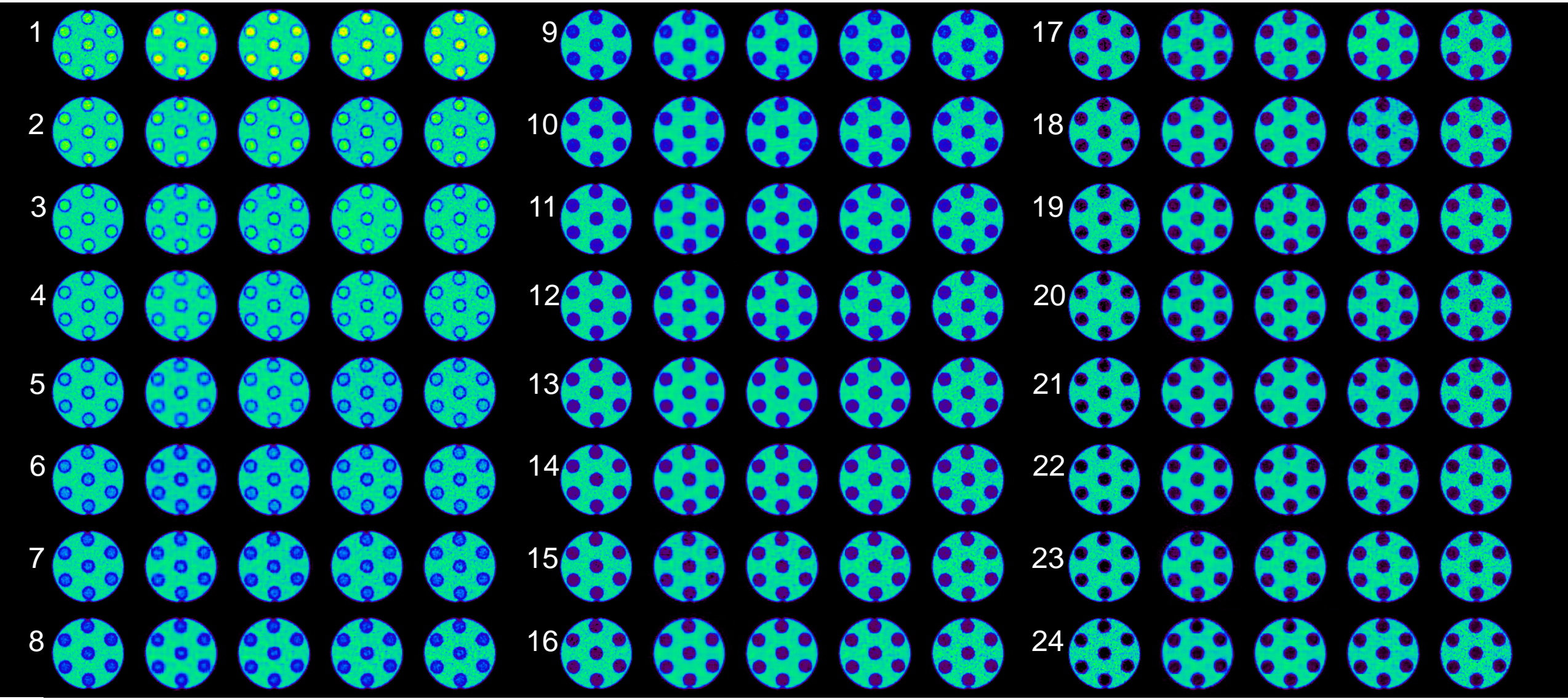

# FigS2(a): OSEM with 3 mm filter

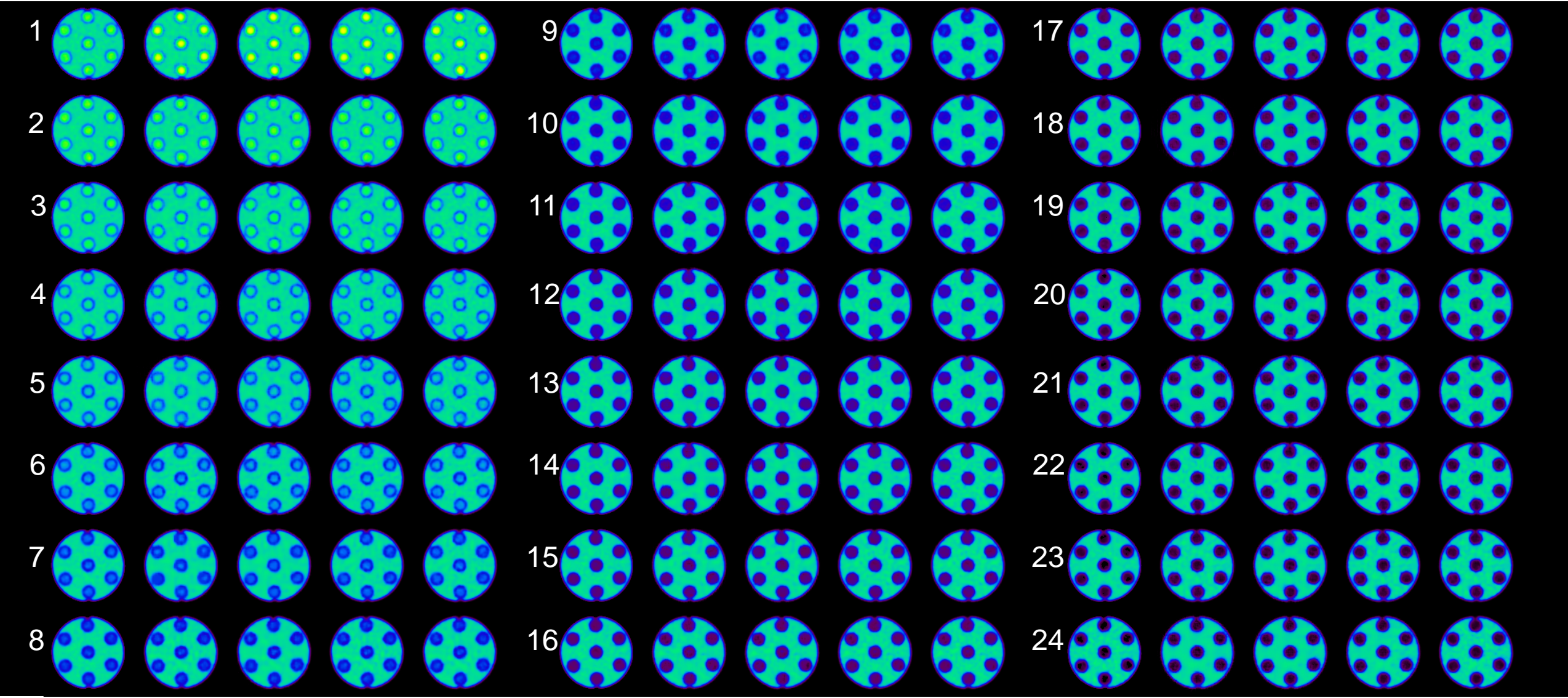

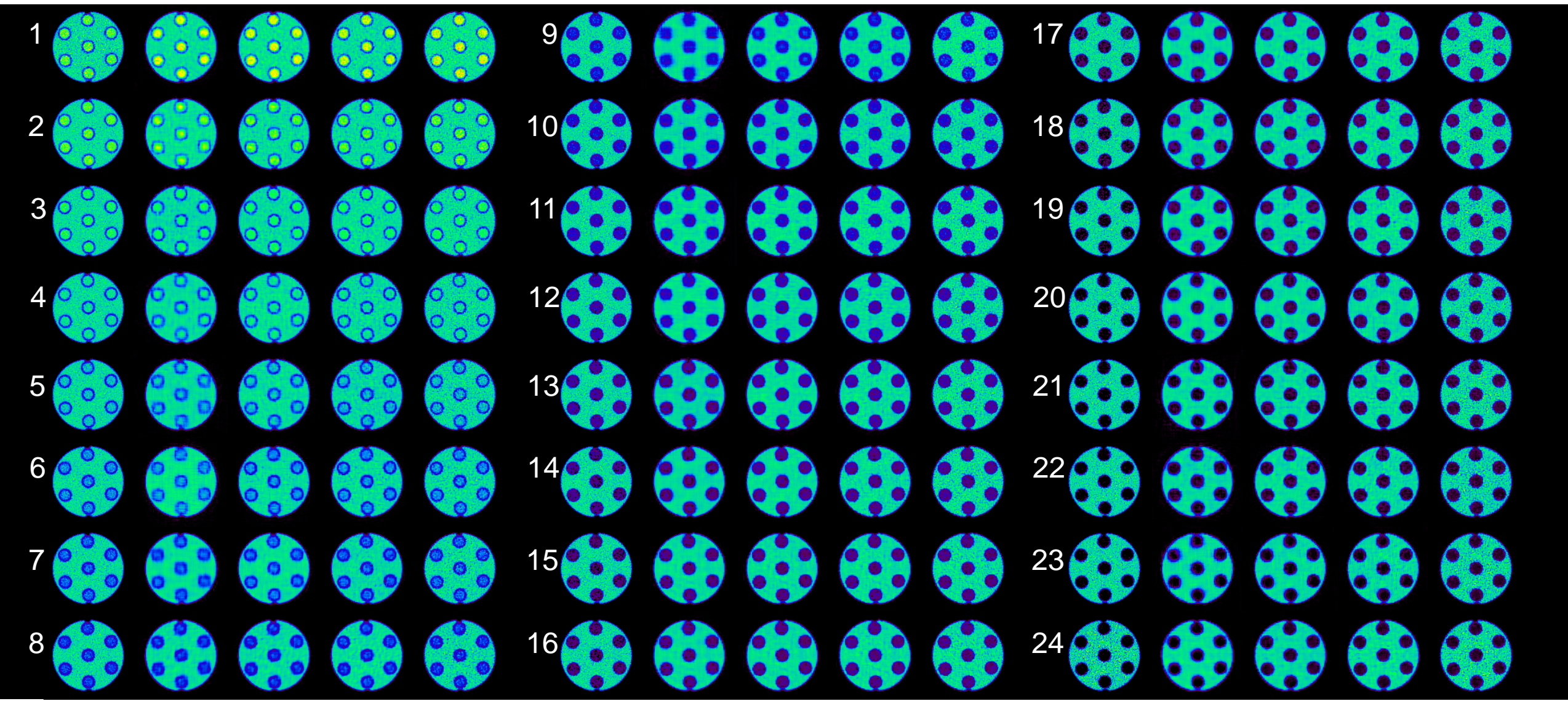

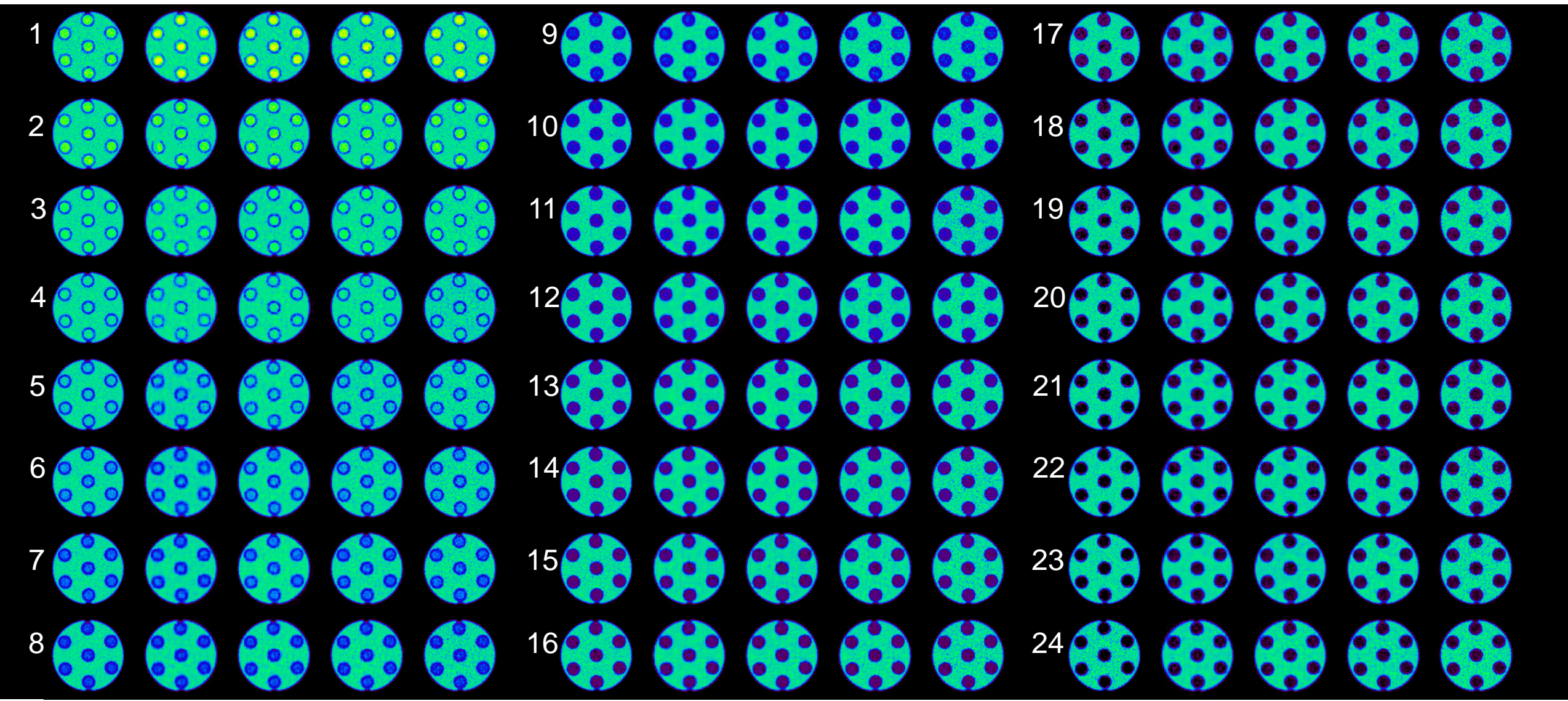

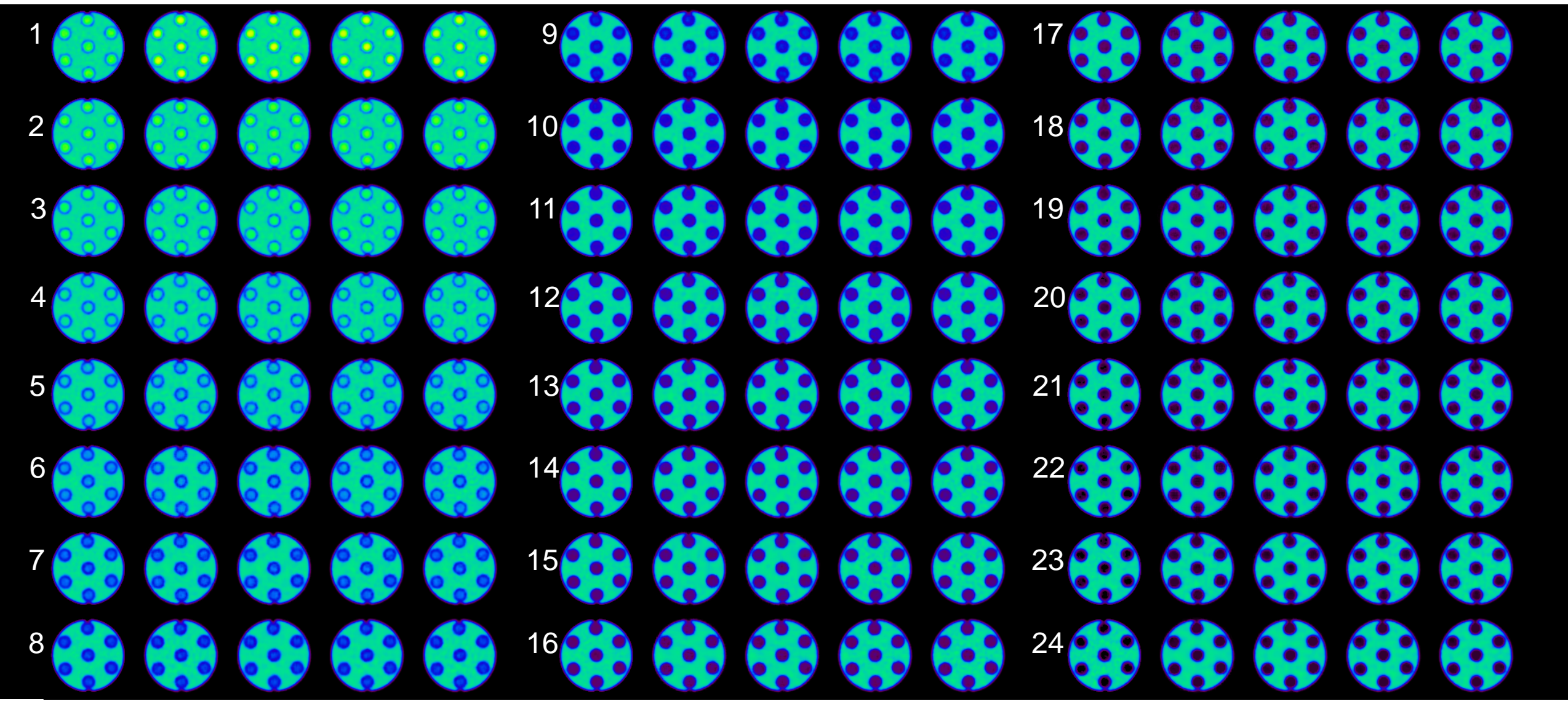